\documentclass[iop]{emulateapj}
\usepackage{apjfonts}
\usepackage{natbib}
\usepackage{graphicx}
\usepackage{times}
\usepackage{xspace} 
\usepackage{booktabs} 
\usepackage{xcolor} % Added by DS 
\usepackage{ulem} % Added by LD, providing the sout command

\slugcomment{\it Submitted --- 2020 December 9}
\shorttitle{GraL VI -- Quad Lenses}
\shortauthors{Stern et al.}
\def\etal{{et al.}}
\def\gaia{{\it Gaia}}
\def\wise{{\it WISE}}

\def\rosat{{\it ROSAT}}

\def\deg{\ifmmode {^{\circ}}\else {$^\circ$}\fi}
\def\kms{\ifmmode {\rm\,km\,s^{-1}}\else
    ${\rm\,km\,s^{-1}}$\fi}
\def\ergcm2s{\ifmmode {\rm\,erg\,cm^{-2}\,s^{-1}}\else
    ${\rm\,erg\,cm^{-2}\,s^{-1}}$\fi}
\def\ergAcm2s{\ifmmode {\rm\,erg\,cm^{-2}\,s^{-1}\,\AA^{-1}}\else
    ${\rm\,erg\,cm^{-2}\,s^{-1}\,\AA^{-1}}$\fi}
\def\ergs{\ifmmode {\rm\,erg\,s^{-1}}\else
    ${\rm\,erg\,s^{-1}}$\fi}
\def\kmsMpc{\ifmmode {\rm\,km\,s^{-1}\,Mpc^{-1}}\else
    ${\rm\,km\,s^{-1}\,Mpc^{-1}}$\fi}

\def\nv{\ion{N}{5} $\lambda$1240}
\def\sio{\ion{Si/O}{4}] $\lambda$1402}

\def\civ{\ion{C}{4} $\lambda$1549}
\def\ciii{\ion{C}{3}] $\lambda$1909}
\def\feii{\ion{Fe}{2} $\lambda \lambda$2343, 2382, 2586, 2599}
\def\mgii{\ion{Mg}{2} $\lambda$2800}
\def\neiiiA{[\ion{Ne}{3}] $\lambda $3343}
\def\nev{[\ion{Ne}{5}] $\lambda $3426}
\def\oii{[\ion{O}{2}] $\lambda $3727}
\def\neiiiB{[\ion{Ne}{3}] $\lambda $3869}
\def\neiiiC{[\ion{Ne}{3}] $\lambda $3968}
\def\oiii{[\ion{O}{3}] $\lambda$5007}

\def\oiiipair{[\ion{O}{3}] $\lambda \lambda$4959,5007}

\def\spose#1{\hbox to 0pt{#1\hss}}
\def\simlt{\mathrel{\spose{\lower 3pt\hbox{$\mathchar"218$}}
     \raise 2.0pt\hbox{$\mathchar"13C$}}}
\def\simgt{\mathrel{\spose{\lower 3pt\hbox{$\mathchar"218$}}
     \raise 2.0pt\hbox{$\mathchar"13E$}}}

\begin{document}

\title{Gaia GraL: Gaia DR2 Gravitational Lens Systems. VI.\\ Spectroscopic Confirmation and Modeling of Quadruply-Imaged Lensed Quasars}

\author{D.~Stern\altaffilmark{1},
S.~G.~Djorgovski\altaffilmark{2},
A.~Krone-Martins\altaffilmark{3, 4}, 
D.~Sluse\altaffilmark{5},
L.~Delchambre\altaffilmark{5},
C.~Ducourant\altaffilmark{6},
R.~Teixeira\altaffilmark{7},
J.~Surdej\altaffilmark{5, 8},
% the rest alphabetical:
C.~Boehm\altaffilmark{9}, 
J.~den~Brok\altaffilmark{10},
D.~Dobie\altaffilmark{11},
A.~Drake\altaffilmark{2}, 
L.~Galluccio\altaffilmark{12}, 
M.~J.~Graham\altaffilmark{2}, 
P.~Jalan\altaffilmark{13, 14},
J.~Kl\"uter\altaffilmark{15, 16},
J.-F.~Le~Campion\altaffilmark{6},
A.~Mahabal\altaffilmark{2},
F.~Mignard\altaffilmark{12},
T.~Murphy\altaffilmark{11},
A.~Nierenberg\altaffilmark{1},
S.~Scarano\altaffilmark{17},
J.~Simon\altaffilmark{1}, 
E.~Slezak\altaffilmark{12}, 
C.~Spindola-Duarte\altaffilmark{7}, \& 
J.~Wambsganss\altaffilmark{15, 18} %\\
% [{\bf AUTHOR LIST ORDER IS TBC}]
}

% ( Christine) : Here is the list of collaborators that should be included in the paper : 
% Alberto Krone-Martins
% Ludovic Delchambre\altaffilmark{}
% Dominique Sluse\altaffilmark{}
% Christine Ducourant\altaffilmark{}
% Ramachrisna Teixeira\altaffilmark{}
% Jean Surdej\altaffilmark{}
% Sergio Scarano\altaffilmark{}
% Laurent Galluccio\altaffilmark{} 
% Jean Francois Le Campion\altaffilmark{}
% Jonas Klueter\altaffilmark{}
% François Mignard\altaffilmark{}
% Eric Slezak\altaffilmark{}
% Céline Boehm\altaffilmark{} 
%%% Ulrich Bastian -- LEAVE OFF
% Joachim Wambsganss\altaffilmark{}

\altaffiltext{1}{Jet Propulsion Laboratory, California Institute of Technology, 4800 Oak Grove Drive, Mail Stop 264-789, Pasadena, CA 91109, USA [e-mail: daniel.k.stern@jpl.nasa.gov]}

\altaffiltext{2}{Cahill Center for Astronomy and Astrophysics, California Institute of Technology, 1216 E.  California Blvd., Pasadena, CA 91125, USA}

\altaffiltext{3}{Donald Bren School of Information and Computer Sciences, University of California, Irvine, Irvine CA 92697, USA}

\altaffiltext{4}{CENTRA, Faculdade de Cincias, Universidade de Lisboa, 1749-016, Lisbon, Portugal}

\altaffiltext{5}{Institut d'Astrophysique et de G\'{e}ophysique, Universit\'{e} de Li\`{e}ge, 19c, All\'{e}e du 6 Ao\^{u}t, B-4000 Li\`{e}ge, Belgium}

\altaffiltext{6}{Laboratoire d’Astrophysique de Bordeaux, Univ. Bordeaux, CNRS, B18N, all\'{e}e Geoffroy Saint-Hilaire, 33615 Pessac, France}

\altaffiltext{7}{Instituto de Astronomia, Geof\'isica e Ci\^encias Atmosf\'ericas, Universidade de S\~{a}o Paulo, Rua do Mat\~{a}o, 1226, Cidade Universit\'aria, 05508-900 S\~{a}o Paulo, SP, Brazil}

\altaffiltext{8}{Astronomical Observatory Institute, Adam Mickiewicz University, Pozna\'n, Poland}

\altaffiltext{9}{School of Physics, The University of Sydney, NSW 2006, Australia}

\altaffiltext{10}{Argelander-Institut für Astronomie, Universität Bonn, Auf dem Hügel 71, D-53121 Bonn, Germany}

\altaffiltext{10}{Sydney Institute for Astronomy, School of Physics, University of Sydney, NSW 2006, Australia}

\altaffiltext{12}{Universit\'{e} C\^{o}te d'Azur, Observatoire de la C\^{o}te d'Azur, CNRS, Laboratoire Lagrange, Boulevard de l'Observatoire, CS 34229, 06304 Nice, France}

\altaffiltext{13}{Aryabhatta Research Institute of Observational Sciences (ARIES), Manora Peak, Nainital 263002, India}
    
\altaffiltext{14}{Department of Physics and Astrophysics, University of Delhi, Delhi 110007, India}

\altaffiltext{15}{Zentrum f\"{u}r Astronomie der Universit\"{a}t Heidelberg, Astronomisches Rechen-Institut, M\"{o}nchhofstr. 12-14, 69120 Heidelberg, Germany}

\altaffiltext{16}{Department of Physics \& Astronomy, Louisiana State University, 261 Nicholson Hall, Tower Dr.,
Baton Rouge, LA 70803-4001, USA}

\altaffiltext{17}{Departamento de F\'{i}sica – CCET, Universidade Federal de Sergipe, Rod. Marechal Rondon s/n, 49.100-000, Jardim Rosa Elze, S\~{a}o Crist\'{o}v\~{a}o, SE, Brazil}

\altaffiltext{18}{International Space Science Institute (ISSI), Hallerstra\ss e 6, 3012 Bern, Switzerland}

\begin{abstract} 

Combining the exquisite angular resolution of \gaia\ with optical light curves and \wise\ photometry, the \gaia\ Gravitational Lenses group (GraL) uses machine learning techniques to identify candidate strongly lensed quasars, and has confirmed over two dozen new strongly lensed quasars from the \gaia\ Data Release 2.  This paper reports on the 12 quadruply-imaged quasars identified by this effort to date, which is a $\sim 20\%$ increase in the total number of confirmed quadruply-imaged quasars. We discuss the candidate selection, spectroscopic follow-up, and lens modeling.  We also report our spectroscopic failures as an aid for future investigations.

\end{abstract}

\keywords{strong gravitational lensing --- quasars}

\section{Introduction}

Strongly lensed quasars provide rare, powerful tools for a range of key studies, including measuring the Hubble constant at intermediate cosmic times \citep[e.g.,][]{Chen:19, Shajib:20}, constraining the properties of dark matter (e.g., Gilman et al. 2019, Nierenberg et al. 2020), inferring the structure near the event horizon of supermassive black holes \citep[SMBHs; e.g.,][]{Pooley:09, Chartas:16}, measuring the accretion disk size of AGN \citep[e.g.,][]{Blackburne:11, Blackburne:15}, constraining the quasar broad emission line region size and geometry \citep[e.g.,][]{Sluse:11, Braibant:17}, measuring black hole spins at redshifts and intrinsic luminosities that would otherwise be inaccessible \citep[e.g.,][]{Walton:15}, and testing general relativity in the strong-gravity regime \citep[e.g.,][]{Collett:18}. A key challenge is finding and confirming these rare sources, particularly the quadruply-imaged ones which are the most constraining for further modelling and physical parameter inference. 

Over the past two years, the \gaia\ Gravitational Lenses working group (GraL) has discovered $\sim 10$\% of all currently confirmed strongly lensed quasars.\footnote{As of December 2020, the Gravitationally Lensed Quasar Database lists 220 confirmed lensed quasars, including 56 quadruply-imaged quasars.} Matching a comprehensive list of known quasars with the \gaia\ source catalog, and then using a supervised machine-learning method to identify which sources were most likely lensed, \citet[][Paper~I]{KroneMartins:18} presented two new quadruply-imaged quasar candidates. \citet[][Paper~II]{Ducourant:18} reported on the \gaia\ Data Release 2 properties of an as-complete-as-possible list of known gravitationally lensed quasars, both confirmed and candidates, that were published prior to \gaia\ Data Release~2 \citep[DR2;][]{Gaia:18}. \citet[][Paper~III]{Delchambre:19} presented a list of 15 new quadruply-imaged quasar candidates based on an update to the supervised machine-learning method used in Paper~I.  \citet[][Paper~IV]{Wertz:19} presented spectroscopic confirmation of a quad lens candidate identified in Papers~I and III, and modeled the lens, including time delays. \citet[][Paper~V]{KroneMartins:20} presents the 15 confirmed doubly-imaged quasars selected by GraL using new candidate selection principles based on unresolved photometric time-series and ground-based images.  This paper (Paper~VI) presents the 12 confirmed quadruply-imaged lensed quasars identified by GraL to date (including the one in Paper~IV, for completeness); we also include one doubly-imaged quasar that was selected as a quad candidate.

New data sets are making this an active time for identifying new lensed quasars, and several other teams are using \gaia\ data, machine learning, and additional approaches to find these rare systems. In a series of papers making use of \gaia\ and \wise\ data, \citet{Lemon:18}, \citet{Lemon:19}, and \citet{Lemon:20} confirmed a total of $\sim 50$ new multiply-imaged quasars, including eight lensed quasars with three or more images.  \citet{Agnello:18} present two confirmed lensed quasars also identified from \gaia. Using supervised machine learning, \citet{Ostrovski:17} identified and confirmed a gravitationally lensed quasar at $z = 2.739$ in the Dark Energy Survey, while \citet{Khramtsov:19} report on machine-learning selection of candidate lensed quasars in the Kilo-Degree Survey. \citet{Ostrovski:18} present the discovery and modeling of a five-image lensed quasar at $z = 3.34$ identified using supervised machine learning from a combination of \gaia\ and ground-based imaging. As one final example from our incomplete survey of the recent literature, \citet{Chao:20} report on a search algorithm for four-image lensed quasars based on their time variability.

This paper is organized as follows.  Section~2 summarizes the GraL lens candidate selection methods. Section~3 summarizes all GraL spectroscopic observations prior to telescope shutdowns caused by COVID-19.  We also include one subsequent observing run from Summer 2020. Section~4 presents the 12 GraL confirmed quaduply-imaged lensed quasars, providing brief details on each system. Section~5 presents those candidates which spectroscopic observations have shown not to be quadruply-imaged lensed quasars, including a discussion of the primary failure modes and details on a few interesting interlopers.  Section~6 presents the lens modeling, and Section~7 summarizes our results.  Throughout this paper, magnitudes are reported in the Vega system.  

% When computing luminosities, we adopt the $\Lambda$CDM concordance cosmology: $H_0 = 70\, {\rm km}\ {\rm s}^{-1}\, {\rm Mpc}^{-1}$, $\Omega_{\rm M} = 0.3$, and $\Omega_\Lambda = 0.7$.

\section{Lens Candidate Selection}

ESA's \gaia\ mission is conducting the largest, most precise, most accurate all-sky astrometric and spectrophotometric survey to date \citep{Gaia:16}. With an effective angular resolution comparable to that provided by {\it Hubble} \citep{Ducourant:18}, \gaia's main goal is to better undestand the Milky Way and produce a three-dimensional dynamical map of our Galaxy based on astrometric measurements of $>10^9$ stars. While achieving this goal, \gaia\ also detects millions of compact galaxies and quasars. This presents a unique opportunity to perform the first homogeneous, magnitude-limited census of strongly lensed quasars over the entire sky, down to lensed image separations of $\sim 0.18^{\prime\prime}$.  Prior to the first data release \citep[DR1;][]{Gaia:16dr1}, \citet{Finet:16} conservatively estimated that \gaia\ would detect $\sim 6.6 \times 10^5$ quasars down to $G < 20$~mag, including $\sim 3000$ resolved, multiply imaged quasars. Considering that \gaia\ is detecting quasars as faint as $G = 20.7$, these estimates are likely too low by $\sim 50$\%.

The GraL team has devised three new methods to identify gravitational lenses \citep[for details, see][]{KroneMartins:18} from \gaia\ Data Release 2 \citep[]{Gaia:18}. The first method relies on a supervised machine learning technique using a Hierarchical Triangular Mesh to identify potential multi-image candidates based on \gaia\ astrometric and photometric properties, and then uses an Extremely Randomized Tree (ERT) to rank these candidates based on simulated lenses. \citet{Delchambre:19} reported this ranking with the ERT probability, $P_{\rm ERT}$.  Note that the ERT probability does not constitute a probability in a mathematical sense, but rather should be viewed as a figure of merit or score that reflects how well matched the source image positions and relative magnitudes are to simulated lens systems. \citet{Delchambre:19} presented 15 candidate quadruply-imaged quasars with $P_{\rm ERT} \geq 0.6$, 11 of which have $P_{\rm ERT} \geq 0.9$.  Of the 21 known, confirmed quadruply-imaged quasars discussed in that paper, 19 have $P_{\rm ERT} \geq 0.6$. Tests indicate a completeness of 77\% for 3-lensed images and 97\% for 4-lensed images (quads), with contamination rates of just 1\% and 0.02\%, respectively.

The second new method uses the information present in the light curves of quasars, supplemented with the \gaia\ astrometry and colors. For an unresolved lensed system (and ignoring microlensing caused by stars in the lensing galaxy), the observed time series is the addition of multiple identical light curves with time delays. Compared to an unlensed quasar, this results in a less stochastic light curve, which we identify using multiscale sample entropy \citep{Ahmed:11} as a proxy for light curve stochasticity.  The optical light curves that we adopted to select the candidates come from the Catalina Real-Time Transient Survey \citep[CRTS;][]{Drake:09} and the Zwicky Transient Facility \citep[ZTF;][]{Bellm:19, Graham:19}.

The third new method uses the information present in the images of quasars, supplemented with the \gaia\ analysis and colors. In brief, the wavelet power spectrum of barely resolved lenses are separated from single quasars, allowing the efficient identification of close systems.  The \gaia\ astrometric and photometric properties of candidates are then combined with mid-IR photometry to efficiently distinguish stellar asterisms from quasars \citep[e.g.,][]{Stern:12, Assef:18}.

%
% TABLE 1:  OBSERVATIONS
% \scriptsize
\begin{deluxetable}{llll}
% \tablewidth{0pt}
\tablecaption{List of observing nights.}
\tablehead{
\colhead{UT Date} &
\colhead{Code} &
\colhead{Telescope/Instr.} &
\colhead{Conditions/Notes}}
\startdata
2018 May 13 & N01-K & Keck/LRIS & photometric \\			% observer: DS
2018 Jun 16 & N02-K & Keck/LRIS & cirrus, poor seeing \\		% observer: SGD
2018 Jul 16 & N03-K & Keck/LRIS & clear, lost 4~hr to elec. prob. \\ % observer: DS
2018 Aug 18 & N04-P & Palomar/DBSP & clouds \\				% observer: JS
2018 Aug 20 & N05-P & Palomar/DBSP & clear, 1\farcs0 seeing \\		% observer: AD
2018 Sep 15 & N06-K & Keck/LRIS & light cirrus, 1\farcs0 seeing \\				% observer: SGD
2018 Oct 03 & N07-K & Keck/LRIS & clear, 0\farcs65 seeing \\		% observer: DS,JdB
2019 Jan 12 & N08-K & Keck/LRIS & clear, good seeing \\			% observer: SGD
2019 Feb 06 & N09-K & Keck/LRIS & clear, good seeing \\			% observer: SGD
2019 Apr 07 & N10-G & Gemini-S/GMOS & queue; clear, 0\farcs5 seeing\\ % observer: queue mode (PI RT)
2019 Apr 07 & N11-N & NTT/EFOSC2 & clear, 0\farcs7 seeing\\ % observer:CD & RT
2019 Apr 08 & N12-N & NTT/EFOSC2 & clear, 0\farcs8 seeing\\ % observer:CD & RT
2019 Apr 09 & N13-N & NTT/EFOSC2 & clear, 1\farcs0 seeing\\ % observer:CD & RT
2019 May 01 & N14-P & Palomar/DBSP & lost to clouds \\			% observer: AD
2019 Jun 01 & N15-K & Keck/LRIS & clear \\  				% observer: SGD
2019 Jun 02 & N16-K & Keck/LRIS & clear; half night \\     		% observer: SGD
2019 Sep 19 & N17-K & Keck/LRIS & cloudy \\		% observer: SGD
2019 Dec 04 & N18-K & Keck/LRIS & cirrus, 0\farcs7 seeing; half night \\ % obs. : SGD
2020 Jan 26 & N19-K & Keck/LRIS & clear, 0\farcs75 seeing \\ % obs. : SGD
2020 Jun 20 & N20-K & Keck/LRIS & scattered cirrus, good seeing % obs. : SGD
\enddata
\label{table:obsevations}
% OBSERVERS:
% - JdB = Jakob den Brok
% - SGD = S.G. Djorgovski
% - AD  = Andrew Drake
% - JS  = Joe Simon
% - DS  = Daniel Stern
% - CD  = Christine Ducourant
% - RT  = Ramachrisna Teixeira
% Note:  All targets from crappy May 2019 Palomar night redone later at Keck.
\end{deluxetable}
% \normalsize

% Keck runs
% 2018may13 - 600/4000, d560, 400/8500, long_1.0
% 2018jun16 - 600/4000, d560, 400/8500, long_1.5
% 2018jul16 - 600/4000, d560, 400/8500, long_1.0
% 2018sep15 - 600/4000, d560, 400/8500, long_1.5
% 2018oct03 - 600/4000, d560, 400/8500, long_1.0
% 2019jan12 - 600/4000, d560, 400/8500, long_1.0
% 2019feb06 - 600/4000, d560, 400/8500, long_1.0
% 2019jun01 - 600/4000, d680, 400/8500, long_1.5
% 2019jun02 - 600/4000, d680, 400/8500, long_1.5
% 2019sep19 - 600/4000, d560, 400/8500, long_1.5
% 2019dec04 - 600/4000, d560, 400/8500, long_1.5
% 2020jan26 - 600/4000, d560, 400/8500, long_1.5
% 2020jun20 - 600/4000, d560, 400/8500, long_1.5

\section{Observations}

Spectroscopic follow-up for the GraL program began in May 2018 and has used the Keck~I telescope atop Maunakea, Hawaii, the 200'' Hale Telescope at Palomar Observatory, California, the 3.6-m New Technology Telescope (NTT) at La Silla, Chile, and the Gemini-South telescope at Cerro Pachon, Chile.  The 20 GraL observing nights are listed in Table~\ref{table:obsevations}, including both dedicated nights (for which we also list nights lost to weather), as well as nights dedicated to other projects but where we were able to obtain at least one GraL spectrum (generally the earlier nights for this program while we were proving our techniques).  In addition to the nights listed in Table~\ref{table:obsevations}, the Gemini-South observations were obtained in queue-mode. The following subsections provide details on these observations, split by telescope/instrument. All observations were processed using standard methods within {\tt IRAF}.  Table~\ref{table:quads} lists the confirmed GraL quadruply-imaged quasars, discussed in detail in \S~4.  The maximum lens separations are based on the astrometry provided in Table~\ref{table:data_for_models} in the Appendix; these values, primarily coming from \gaia, are also reported in the text of \S4. Table~\ref{table:asterisms} in the Appendix lists the spectroscopically observed GraL targets which proved not to be lensed quasars, and are discussed in \S~5.

% Ludovic Delchambre suggested added reference:  \com{LD}{Add a reference here: e.g. https://ui.adsabs.harvard.edu/abs/1986SPIE..627..733T/abstract}

\subsection{Keck/LRIS}

The primary instrument for GraL spectroscopic follow-up has been the Low Resolution Imaging Spectrometer \citep[LRIS;][]{Oke:95} on the Keck~I telescope, used for 13 of the 20 GraL nights listed in Table~\ref{table:obsevations}. For all listed LRIS observing runs, we used the 600 $\ell$~mm$^{-1}$ blue grism ($\lambda_{\rm blaze} = 4000$~\AA), and the 400 $\ell$~mm$^{-1}$ red grating ($\lambda_{\rm blaze} = 8500$~\AA). Most runs used the 5600~\AA\, dichroic, though the 6800~\AA\, dichroic was used for the June 2019 nights.  The slit widths were tailored to the observing conditions for each night, with the 1\farcs5 used for most nights, though we used the 1\farcs0 slit on the May 2018, July 2018, October 2018, January 2019, and February 2019 nights. Our LRIS instrument configuration covers the full optical window at moderate resolving power, $R \equiv \lambda / \Delta \lambda \approx 1000$ for the 1\farcs5 slit and $R \approx 1500$ for the 1\farcs0 slit (for objects filling the slit). Over the course of these nights, standard stars from \cite{Massey:90} were observed for flux calibration.

\subsection{Palomar/DBSP}

Several lens candidates were observed with the Palomar Double Spectrograph (DBSP) in  August 2018, and a Palomar night dedicated for this project was lost to clouds in May 2019.  None of the confirmed quadruply-imaged quasars were observed with Palomar, though two doubly-imaged quasars in \citet{KroneMartins:20} were confirmed on these runs, and several of the GraL lens candidates were invalidated at Palomar (Table~\ref{table:asterisms}). We used the 1\farcs0 slit, the 5600\AA\ dichroic, the 600~$\ell$~mm$^{-1}$ blue grism ($\lambda_{\rm blaze} = 4000$~\AA), and the 400~$\ell$~mm$^{-1}$ red grating ($\lambda_{\rm blaze} = 8500$~\AA) for all Palomar observations reported here.  This instrument configuration covers the full optical window at moderate resolving power, $R \approx 1250$. Standard stars from \cite{Massey:90} were observed for flux calibration.

\subsection{NTT/EFOSC2}

We used the ESO Faint Object Spectrograph and Camera (v.2; EFOSC2) on the NTT telescope on the nights of UT 2019 April 7-9 (PI: Ducourant; Program ID 0103.A-0077). Conditions were clear with $\sim 1\arcsec$ seeing. We observed 9 candidates using the 5\farcs0 and 1\farcs5 width slits, Grism1 covering 3185-10940~\AA~ ($\lambda_{\rm blaze} = 4500$~\AA), and the GG375 order-blocking filter. The LTT3864 and LTT7379 spectrophotometric standards from \citet{Hamuy:94} were observed on the first two nights for flux calibration.

\subsection{Gemini-S/GMOS-IFU}

Seven candidates were observed in queue mode using the Gemini-South Multi-Object Spectrograph \citep[GMOS;][]{Hook:04} in integral field unit \citep[IFU;][]{Allington-Smith:02} mode (PI: Teixeira; Program ID GS-2019A-Q-104). We used the lowest resolution grating available at the time of these observations, namely the 400 $\ell\, {\rm mm}^{-1}$ grating which covered the $7000 - 8200$~\AA\, spectral range with a resolving power $R \sim 2000$.  We used the IFU in two-slit mode in order to fit the targets within the IFU field ($7 \times 5$ arcsec$^2$) and obtained a single 1200~s exposure of each target.  The observational conditions varied between targets, though all occurred with cloud cover below 70\%, image quality better the 70\% (i.e., seeing better than 0\farcs75), and airmass below 1.5.  The spectra were flux calibrated using standards supplied by the Gemini baseline calibrations. The limited spectral range prevented clear assessments for most targets; only the target observed on UT 2019 April 7 proved useful and is discussed below.

% According the Gemini Integration Time Calculator these configurations could cover important spectral lines for possible quasars and the deflectors, like MgII, NeVI, NeIII, OIII, Hgama  for redshifts inside a range of 0.8 – 1.5. 

%\subsection{SOAR/XXX}
%
%[{\it Were there any SOAR observations?  No.}]

%
% FIGURE 1 - QUAD IMAGE STAMPS
\begin{figure*}[!ht]
\begin{center}
\includegraphics[width=1.0\textwidth]{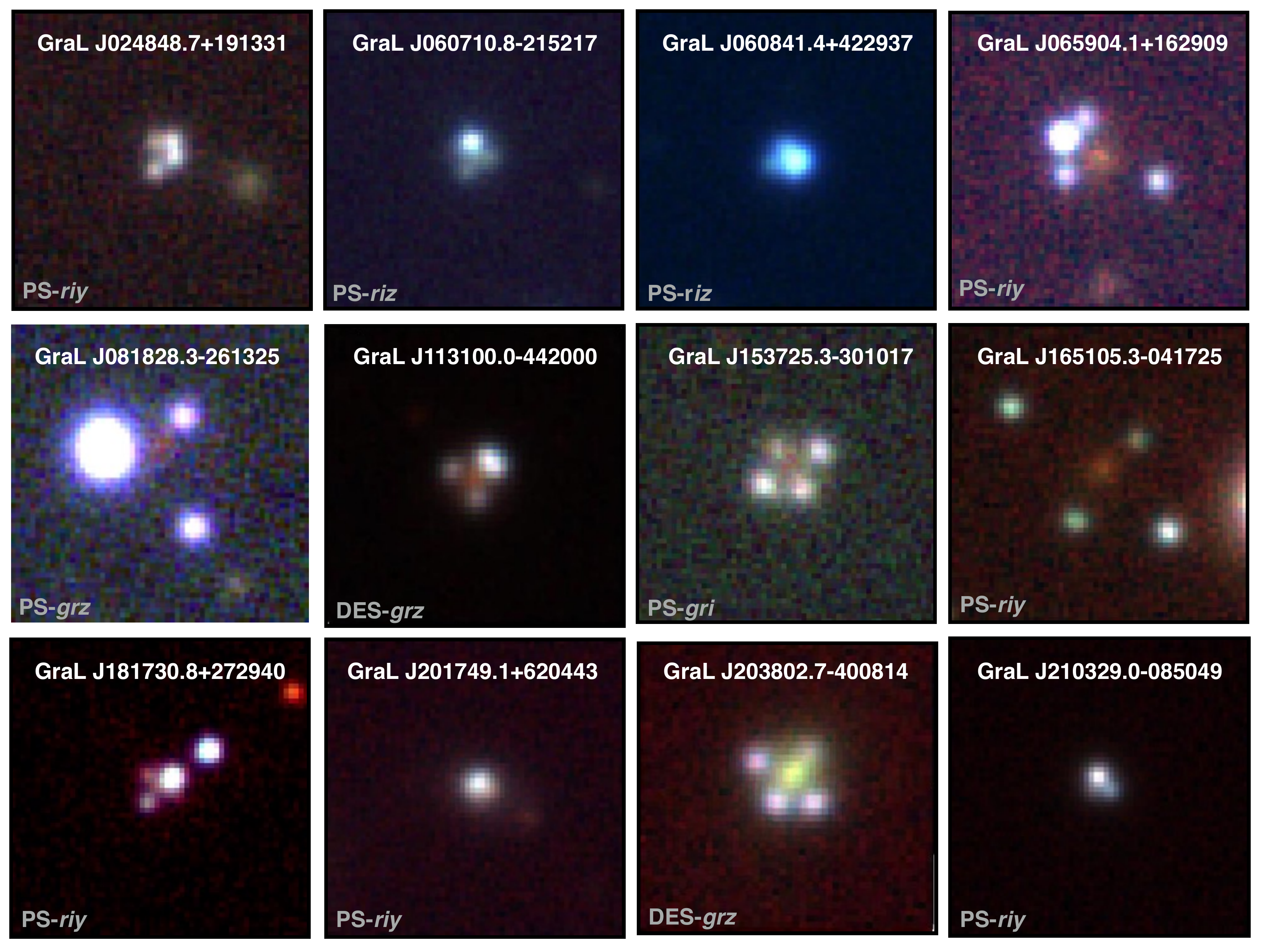} 
\caption{False-color images, 15\arcsec\ on a side, of the confirmed quadruply-imaged quasars.  North is up, and east is to the left. Bottom left corner of each image indicates whether the imaging is from PanSTARRS (PS) or the Dark Energy Survey (DES), and which filters were mapped to the RGB colors.
\label{fig:images}}
\end{center}
\end{figure*}

%
% TABLE 2:  CONFIRMED QUAD LENSES
% \scriptsize
\begin{deluxetable*}{lcccccl}
% \tablewidth{0pt}
\tablecaption{Confirmed GraL quad lenses.}
\tablehead{
\colhead{} &
\colhead{Max.} &
\colhead{} &
\colhead{} &
\colhead{Exposure} &
\colhead{} &
\colhead{}\\
\colhead{Name} &
\colhead{Sep'n} &
\colhead{Night} &
\colhead{PA} &
\colhead{Time (s)} &
\colhead{$z$} &
\colhead{Notes}}
\startdata
GraL J024848.7+191331   & 1\farcs7 & N03-K & $-10\deg$ &          450 & 2.424 & Paper III \\
GraL J060710.8$-$215217 & 1\farcs7 & N19-K & 20$\deg$ & $2\times400$ & 1.302 & \\
                  & & N19-K & 352$\deg$ & $2\times500$ & & \\
                  & & N19-K & 57$\deg$ & $2\times500$ & & \\
GraL J060841.4+422937   & 1\farcs3 & N18-K & 222$\deg$ & $2\times300$ & 2.345 & ``Auriga's Slingshot''\\
                &  & N18-K & 275$\deg$ & $2\times300$ & & \\
GraL J065904.1+162909   & 6\farcs8 & N06-K &  66$\deg$ & $2\times300$ & 3.083 & Paper III; ``Orion's Crossbow'' \\
        & & N07-K &   0$\deg$ &          600 & & \\
        & & N07-K &  40$\deg$ &          600 & & \\
        & & N08-K &  66$\deg$ & $2\times300$ & & deflector at $z=0.766$ \\
 		& & N08-K &   5$\deg$ & $2\times300$ & & \\
 		& & N08-K & 132$\deg$ & $2\times300$ & & \\
GraL J081828.3$-$261325 & 6\farcs2 & N09-K &   0$\deg$ & $3\times300$ & 2.164 & ``Argo's Rose'' \\
GraL J113100.0$-$442000 & 1\farcs6 & N01-K &  60$\deg$ &          300 & 1.090 & Paper I,III,IV; ``Centaurus' Victory'' \\
        & & N01-K & 135$\deg$ &          300 & & \\
        & & N10-G &  10$\deg$ &         1200 & & IFU \\
GraL J153725.3$-$301017 & 3\farcs3 & N03-K & 165$\deg$ & $2\times600$ & 1.721 & Paper III; ``Wolf's Paw'' \\
GraL J165105.3$-$041725 & 10\farcs1 & N13-N &  51$\deg$ &       1800 & 1.451 & ``Dragon Kite''\\
        & & N13-N & 142$\deg$ &         1800 & & \\
        & & N15-K &  51$\deg$ & $2\times600$ & & \\
	    & & N15-K & 142$\deg$ & $2\times600$ & & deflector at $z=0.591$ \\
GraL J181730.8+272940 & 1\farcs8 & N03-K & 135$\deg$ &          300 & 3.074 & Paper III; ``Hercules' Sword'' \\
GraL J201749.1+620443   & 0\farcs7 & N03-K & 120$\deg$ &          300 & 1.724 & Paper III\\
GraL J203802.7$-$400814 & 2\farcs5 & N01-K &  90$\deg$ &          300 & 0.775 & Paper I,III, \citet{Agnello:18}; deflector at $z=0.228$; ``Microscope Lens''\\
GraL J210329.0$-$085049 & 1\farcs0 & N15-K &   0$\deg$ & $2\times300$ & 2.455 & ``Aquarius' Tear'' 
\enddata
\label{table:quads}
\tablecomments{In addition to these 12 quadruply-imaged lensed quasars, we also confirm the doubly-imaged lensed quasar GraL~J201454.2$-$302452 (see \S~5).}
\end{deluxetable*}
% \normalsize

% name suggestions from Laurent Galluccio:
% GraL J081830.5+060137.9:  Hydra's Emerald
% GraL J153725.3-301017: Wolf footprint
% GraL J024612.2-184505.1: Eridanus keyhole
% GraL J181730.8+271940: Hercules sword
% GraL J203802.7-400814: Microscope Lens
% GraL J060841.4+422937: Auriga engagement ring
% GraL J210329.0-085049: Aquarius teardrops

\section{Confirmed Quad Lenses}

In the following, we present details on all quadruply-imaged quasars confirmed by the GraL collaboration to date. Figure~1 presents false-color optical images of these sources, and Figures~2-4 present their processed spectra.

{\bf GraL J024848.7+191331 ---}  This source was presented in \cite{Delchambre:19} as a candidate lensed quasar with three \gaia-detected lensed images, an ERT probability of $P_{\rm ERT} = 0.88$, and a maximum separation of 1\farcs7.  The PanSTARRS image shows all four components.  A single Keck spectrum, aligned to observe the two brightest components, confirms the lensed nature of the source with two spatially separated, nearly identical spectra of the source broad absorption line (BAL) quasar.  The spectra show strong \ion{O}{6}~$\lambda\lambda 1032.0, 1037.6$ emission, as well as a foreground absorption line system at $z = 1.037$, potentially associated with the lensing galaxy.  There are also strong, broadened unidentified absorption features at observed wavelengths of 4525 and 6292~\AA, corresponding to rest-frame 1321 and 1838~\AA\, respectively, at the quasar redshift. No radio or X-ray counterparts are reported by NED.

% W1 = 14.328 pm 0.028
% W2 = 13.014 pm 0.027
% W3 =  9.304 pm 0.039
% W4 =  7.029 pm 0.091

\begin{figure*}[!htp]
\begin{center}
\includegraphics[width=1.8\columnwidth]{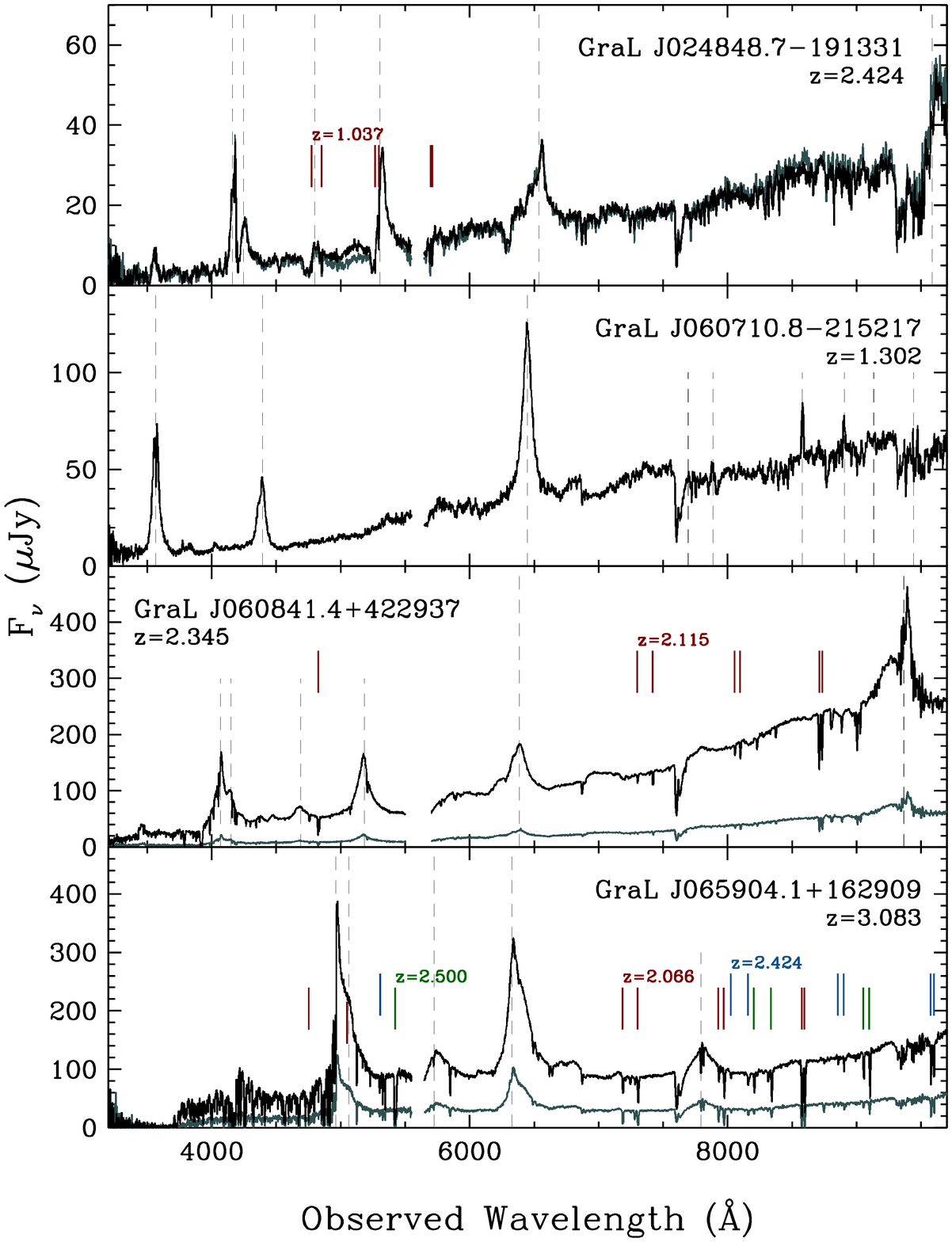}
\caption{Spectra of the first four sources presented herein.  Key emission lines are indicated with vertical dashed lines (in order, Ly$\alpha$, \nv, \sio, \civ, \ciii, \mgii, \neiiiA, \nev, \oii, \neiiiB, \neiiiC, H$\gamma$, H$\delta$, H$\beta$, and \oiiipair; given the range of redshifts, only a subset of emission lines are indicated for each source). Prominent absorption lines are presented with short, solid, vertical lines.  Labeled systems show absorption from \civ, \feii, and \mgii\ (not all lines are indicated or detected for each system).  See text for details.
\label{fig:spectra1}}
\end{center}
\end{figure*}

{\bf GraL J060710.8$-$215217 ---} This candidate was selected from a more recent ERT run using \gaia\ DR2, and has a compact, 1\farcs7 separated configuration with \wise\ colors indicative of the presence of an AGN.  The lensed quasar nature was confirmed in January 2020 based on Keck spectroscopy at multiple position angles (PAs) that all clearly showed traces of two very close quasars at the same redshift.  Figure~\ref{fig:spectra1} presents the spectrum obtained at PA=20\deg, obtained with a single 3\farcs0 wide extraction aperture. No radio or X-ray counterparts are reported by NED.

{\bf GraL J060841.4+422937 ---}  This candidate was also selected from the more recent ERT run using \gaia\ DR2, and has a compact, 1\farcs3 separated configuration.  The source had a high (86\%) likelihood of being a quasar from the Million Quasars Catalog \citep[ver.~6.4c;][]{Flesch:15}.  Keck spectroscopy was obtained in December 2019 at two PAs, both of which clearly show two spatially distinct spectra of the same lensed quasar.  Figure~\ref{fig:spectra1} presents the spectra obtained at PA=275\deg.  An absorption system with multiple features from \civ\ to \mgii\ is detected at $z = 2.112$, which is presumed too close to the quasar redshift to be associated with the lensing galaxy. Located in the constellation Auriga, the charioteer, this source has been given the name ``Auriga's Slingshot''. No radio or X-ray counterparts are reported by NED.

{\bf GraL~J065904.1+162909 ---}  This source was presented in \cite{Delchambre:19} as a candidate lensed quasar with three \gaia-detected lensed images, an ERT probability of $P_{\rm ERT} = 0.94$, and a wide maximum separation of 5\farcs2; including SDSS imaging which detects a fourth image, the maximum separation increases to 6\farcs8.  The PanSTARRS image shows all four components, in a kite-like pattern of blue sources around the clearly detected lens galaxy with red optical colors.  Owing to this configuration and its location on the sky, we have dubbed this source ``Orion's Crossbow''.  We obtained several Keck spectra over multiple observing runs at multiple PAs, and confirmed that all four blue components correspond to a quasar at $z = 3.083$; Figure~2 presents just the two spectra of the most separated components, obtained in January 2019 at PA $= 66\deg$.  That spectrum also confirmed the lensing galaxy to be an early-type galaxy at $z = 0.766$ with strong absorption from Ca~HK and Mg~Ib, as well as a pronounced 4000~\AA\ break.  The quasar spectra show multiple absorption line systems, at redshifts of  $z = 2.066$, 2.424, and 2.500. Absorption lines from the lensing galaxy are not seen, likely due to its low redshift placing key features in the dense Ly$\alpha$ forest. No radio or X-ray counterparts are reported by NED.

% W1 = 13.842 pm 0.027
% W2 = 13.247 pm 0.030
% W3 = 10.100 pm 0.073
% W4 =  7.497 pm 0.155

{\bf GraL J081828.3$-$261325 ---}  This is a quadruply-imaged quasar at $z = 2.164$.  Two components are closely spaced, with the other two a few arcseconds to the west.  Faint red flux potentially associated with the lensing galaxy is evident in Figure~1.  This source was identified from a revised ERT search for lensed quasars performed in 2019 combined with the wavelet method discussed in \S~2.  The Keck observation, presented in Figure~2, targeted the compact pair and shows a spatially extended system in the two-dimensional spectrum, indicative of a lensed quasar, though the seeing was insufficient to resolve the individual components.  Instead, a single wide extraction was used. The quasar itself has a red continuum and weak Ly$\alpha$ emission, indicative of moderate absorption either at the quasar itself or along the line of sight.  We identify several absorption lines due to a foreground system at $z = 2.088$, unlikely to be associated with the lensing galaxy given the small redshift difference between the quasar and this absorption line system.  No radio or X-ray emission are reported by NED.  With a \wise\ $22~\mu$m magnitude of $W4 = 4.82 \pm 0.03$, this is the brightest mid-IR lensed quasar presented here.  Based on it's red colors and location in the Puppis constellation,  which references the stern deck of the Argo, the ship on which Jason and the Argonauts sailed to get the Golden Fleece, we have named this source ``Argo's Rose''.

% W1 = 12.618 pm 0.024
% W2 = 11.301 pm 0.020
% W3 =  7.382 pm 0.017
% W4 =  4.823 pm 0.030

% other suggested name for 0818-2613:  Hydra's Emerald

{\bf GraL J113100.0$-$442000 ---}  This is the quadruply-imaged quasar presented as a candidate in \citet{KroneMartins:18} and \citet{Delchambre:19}, and then spectroscopically confirmed and modelled in \citet{Wertz:19}. The latter is a detailed paper dedicated to this single source.  This was the first confirmed gravitational lens to be discovered from a machine learning technique that relied only on the relative positions and fluxes of the observed images, without consideration of color information, and thus earned the name ``Centaurus' Victory.'' \citet{Delchambre:19} reported an ERT probability of $P_{\rm ERT} = 0.96$, and a maximum separation of 1\farcs6.  The source is at intermediate redshift, $z = 1.090$, and the Keck spectroscopy shows no evidence of the central lens galaxy.  Gemini IFU observations separated the two quasar images with the smallest angular separation, called A and B in \citet{Wertz:19}, as seen in Fig.~3. The lensed quasar is likely associated with 1RXS~J113058.9-441949 from the {\it ROSAT} all-sky survey bright source catalogue \citep{Voges:99}, which reports an X-ray astrometric uncertainty of 32\farcs5 in each coordinate. No radio emission is reported by NED.

% W1 = 13.577 pm 0.025
% W2 = 12.499 pm 0.024
% W3 =  9.340 pm 0.032
% W4 =  6.527 pm 0.060

{\bf GraL J153725.3$-$301017 ---} This source was presented in \cite{Delchambre:19} as a candidate lensed quasar with three \gaia-detected lensed images, an ERT probability of $P_{\rm ERT} = 0.97$, and a wide maximum separation of 3\farcs3.  The PanSTARRS image shows all four components in a classic diamond configuration, with hints of a red lensing galaxy in the center.  We obtained two Keck spectra at PA=165\deg, simultaneously observing the eastern and western pair of lensed quasar images in turn. We confirmed that all four blue components correspond to a quasar at $z = 1.721$; Figure~3 presents the spectra of the two western components, which show an AGN with strong narrow-line components superposed on broad emission lines. There was no evidence of the lensing galaxy in these spectra, nor do we detect a foreground absorption line system. Located in the constellation Lupus with a configuration suggestive of a paw print, this source has earned the moniker ``The Wolf's Paw''.  No radio or X-ray counterparts are reported by NED.

% W1 = 14.599 pm 0.035
% W2 = 13.901 pm 0.051
% W3 = 11.364 pm 0.200
% W4 =  8.586 pm 0.422 (i.e., 2.6-sigma)

{\bf GraL J165105.3$-$041725 ---} This is the widest separation quadruply-imaged quasar in our sample, with a maximum separation of 10\farcs1.  It was identified from an ERT analysis, initially confirmed at NTT and then deeper spectra were obtained with Keck with two longslit spectra aligned to observe both pairs of lensed quasar images.  All four spectra show a quasar at $z = 1.451$; Figure~3 presents the Keck/LRIS spectra for the two components observed with the PA = 142\deg\ slit.  That spectrum also confirmed the lensing galaxy to be an early-type galaxy at $z = 0.591$ with strong absorption from Ca~HK and Mg~Ib, as well as a pronounced 4000~\AA\ break.  No absorption from the lensing galaxy is evident in the quasar spectra, though an absorption line system at $z = 1.375$ with strong \civ\ and \mgii\ absorption is evident.  Owing to the large size of this diamond-shaped lens system and its location within the Ophiuchus, or Serpent Bearer, constellation, this lens has been named the ``Dragon Kite''.  No radio or X-ray counterparts are reported by NED.

% W1 = 13.864 pm 0.027 + 12.627 pm 0.024
% W2 = 13.095 pm 0.031 + 12.574 pm 0.027
% W3 = 10.636 pm 0.117 + 12.066 pm 0.454 (2.4 sigma)
% W4 =  8.516 pm 0.353 + not detected 

{\bf GraL J181730.8+272940 ---} This source was presented in \cite{Delchambre:19} as a candidate lensed quasar with three \gaia- detected lensed images, an ERT probability of $P_{\rm ERT} = 0.91$, and a maximum separation of 1\farcs8.  Our Keck spectrum at PA=135\deg\ confirmed that the fainter SE component and the brighter NW component of the compact configuration are a lensed quasar at $z = 3.074$ with strong \civ\ BAL features and weak Ly$\alpha$.  The bright point source 2\arcsec\ NW of the lens was also observed in that slit configuration and shown to be a Galactic mid-type star with absorption from Ca~HK, H$\alpha$, and the calcium triplet.  Owing to the configuration suggestive of a hilt and a blade, we have named this source ``Hercules' Sword''.  The faint radio source NVSS~J181731+272951 ($S_{\rm 1.4~GHz} = 3.2$~mJy) is approximately 10\arcsec\ NE of the lens, but is unlikely associated.  No X-ray counterparts are reported by NED.

% W1 = 14.127 pm 0.028
% W2 = 13.293 pm 0.028
% W3 =  9.530 pm 0.036
% W4 =  6.643 pm 0.063

%{\bf GraL~J2014$-$3024.} Previously reported in %\citet{Delchambre:19}.
% W1 = 15.557 pm 0.049
% W2 = 14.572 pm 0.065
% W3 = 11.064 pm 0.155
% W4 - not detected

{\bf GraL J201749.1+620443 ---} This source was presented in \cite{Delchambre:19} as a candidate lensed quasar with three \gaia-detected lensed images, an ERT probability of $P_{\rm ERT} = 0.74$, and a compact maximum separation of just 0\farcs7.  This was the smallest lens system candidate presented in that paper, as well as lowest ERT probability that we have confirmed; only one candidate had an even lower $P_{\rm ERT} = 0.60$ in that paper, which we spectroscopically identified as an asterism of Galactic stars (J011559.5+562506; Table~6).  The Keck spectrum did not clearly resolve this quasar as spatially extended, which is consistent with the compact lens configuration, comparable to the $\sim 1\farcs0$ seeing of that night. The spectrum itself shows a reddened BAL quasar with several unidentified narrow absorption features (e.g., at 4737, 4778, and 7090 \AA). No radio or X-ray counterparts are reported by NED.  

% Delchambre+19 max separation was 0.9" [TBC]

% W1 = 13.862 pm 0.025
% W2 = 12.843 pm 0.023
% W3 =  9.935 pm 0.038
% W4 =  7.545 pm 0.097

{\bf GraL J203802.7$-$400814 ---} This lens, also known as WGD2038-4008, was previously reported in \citet{Agnello:18} from a combined search in \wise\ and \gaia\ DR1 over the Dark Energy Survey (DES) footprint, and was independently co-discovered from the GraL ERT analysis and reported in \citet{KroneMartins:18} and \citet{Delchambre:19}. The latter GraL paper reports it as having four \gaia-detected lensed images, an ERT probability of $P_{\rm ERT} = 1.00$, and a maximum separation of 2\farcs5.  The Keck observation, obtained at a suboptimal PA=90\deg\, shows two spatially distinct spectra of the reddened lensed quasar, as well as the lensing host galaxy.  The \oiii\ emission line from the lensed quasar is slightly physically extended, and the lensing galaxy at $z = 0.228$ is clearly identified with \oii\ emission superposed on an early-type galaxy with absorption due to Ca~HK, the G-band, and Mg~Ib (all indicated with vertical solid marks in Figure~4).  Owing to it's location in the Microscopium constellation, we have named this source the ``Microscope Lens''.  This lensed system is associated with the \rosat\ X-ray source 1RXS~J203801.8$-$400818 \citep{Voges:99}. No radio counterpart is identified by NED.

% Delchambre+19 max separation was 2.9" [TBC]

% W1 = 11.432 pm 0.022
% W2 = 10.246 pm 0.020
% W3 =  7.489 pm 0.017
% W4 =  5.112 pm 0.037

{\bf GraL J210329.0$-$085049 ---} This lensed quasar was identified as a candidate from a \gaia\ ERT analysis subsequent to \citet{Delchambre:19}, with a compact maximum separation of 1\farcs0.  Two lensed quasar spectra are clearly identified in the blue arm of the LRIS data, though they merge at longer wavelengths.  The spectra of this source presented in Figure~4 show the two blue arms (extracted with 0\farcs5 apertures), but a single, scaled red spectrum (extracted with a 2\farcs0 aperture).  We identify a foreground absorption line system at $z = 0.768$, potentially associated with the lensing galaxy.  Similar to the doubly-imaged quasar GraL~J234330.6+043557.9 discussed in \citet{KroneMartins:20}, the absorption is significantly stronger in one lensed image (see Fig.~4) than the other.  It is assumed that this foreground absorption corresponds to the lensing galaxy, with one of the lensed quasar images intercepting a more dense region of the lensing galaxy. Because of its compact configuration and location on the celestial sphere, we've named this source ``Aquarius' Tear''.  This system is detected as the X-ray source 1RXS~210328.9$-$085039 \citep{Voges:99}, and no radio counterpart is identified by NED.

% Delchambre+19 max separation was 0.8" [TBC]

% W1 = 14.300 pm 0.031
% W2 = 13.199 pm 0.032
% W3 =  9.654 pm 0.079
% W4 =  7.769 pm 0.322

%
% FIGURE 2:  SPECTROSCOPY, PART 1/3

%
% FIGURE 3:  SPECTROSCOPY, PART 2/3
\begin{figure*}[!htp]
\begin{center}
\includegraphics[width=1.8\columnwidth]{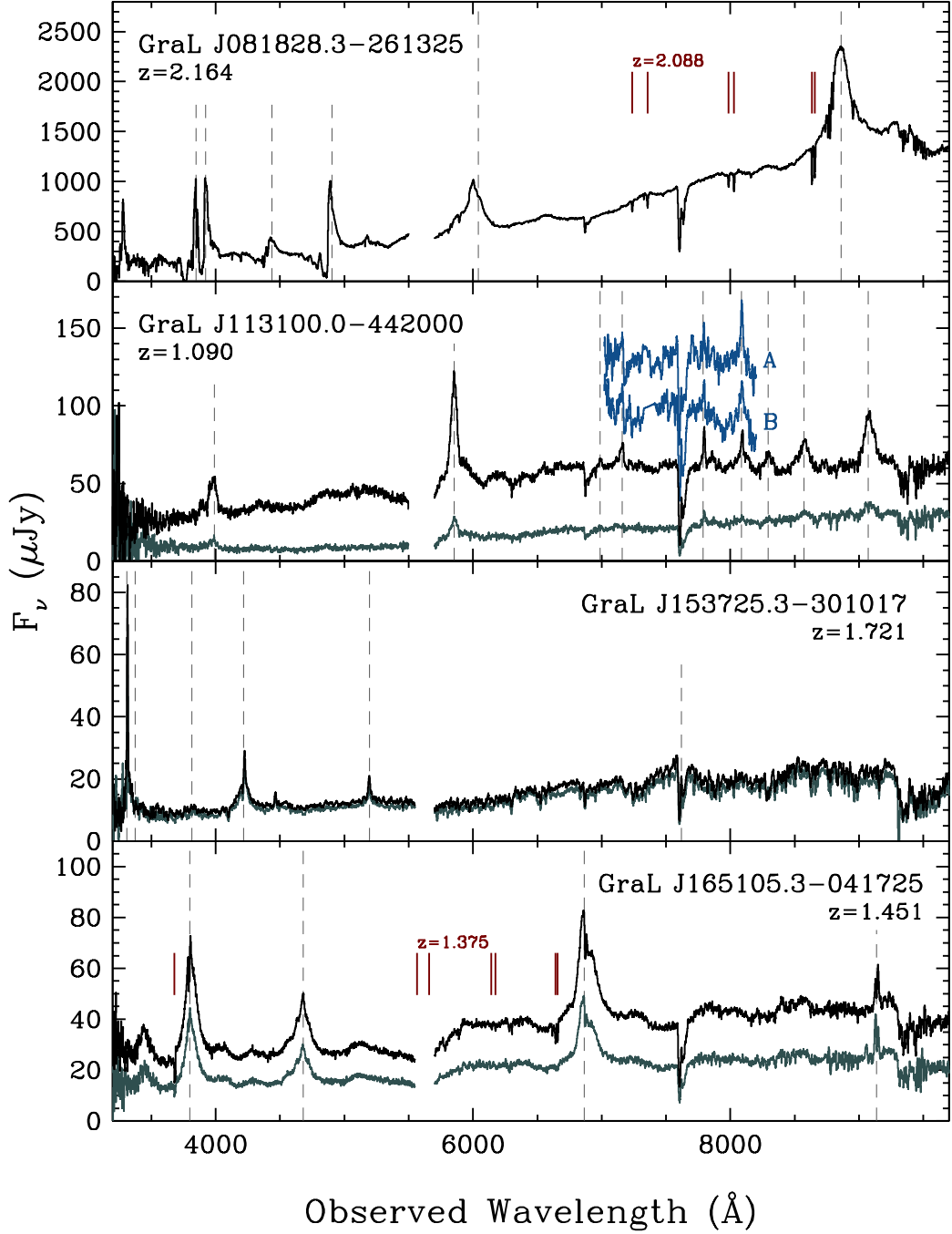}
\caption{Spectra of the second four confirmed quadruply-imaged GraL quasars, as per Figure~2.  For GraL~1131100.0$-$442000, the shorter, offset blue spectra are the individual, deblended A and B components from the GMOS-S/IFU observations, scaled to make more visible (see text for details). 
\label{fig:spectra2}}
\end{center}
\end{figure*}

%
% FIGURE 4:  SPECTROSCOPY, PART 3/3
\begin{figure*}[!htp]
\begin{center}
\includegraphics[width=1.8\columnwidth]{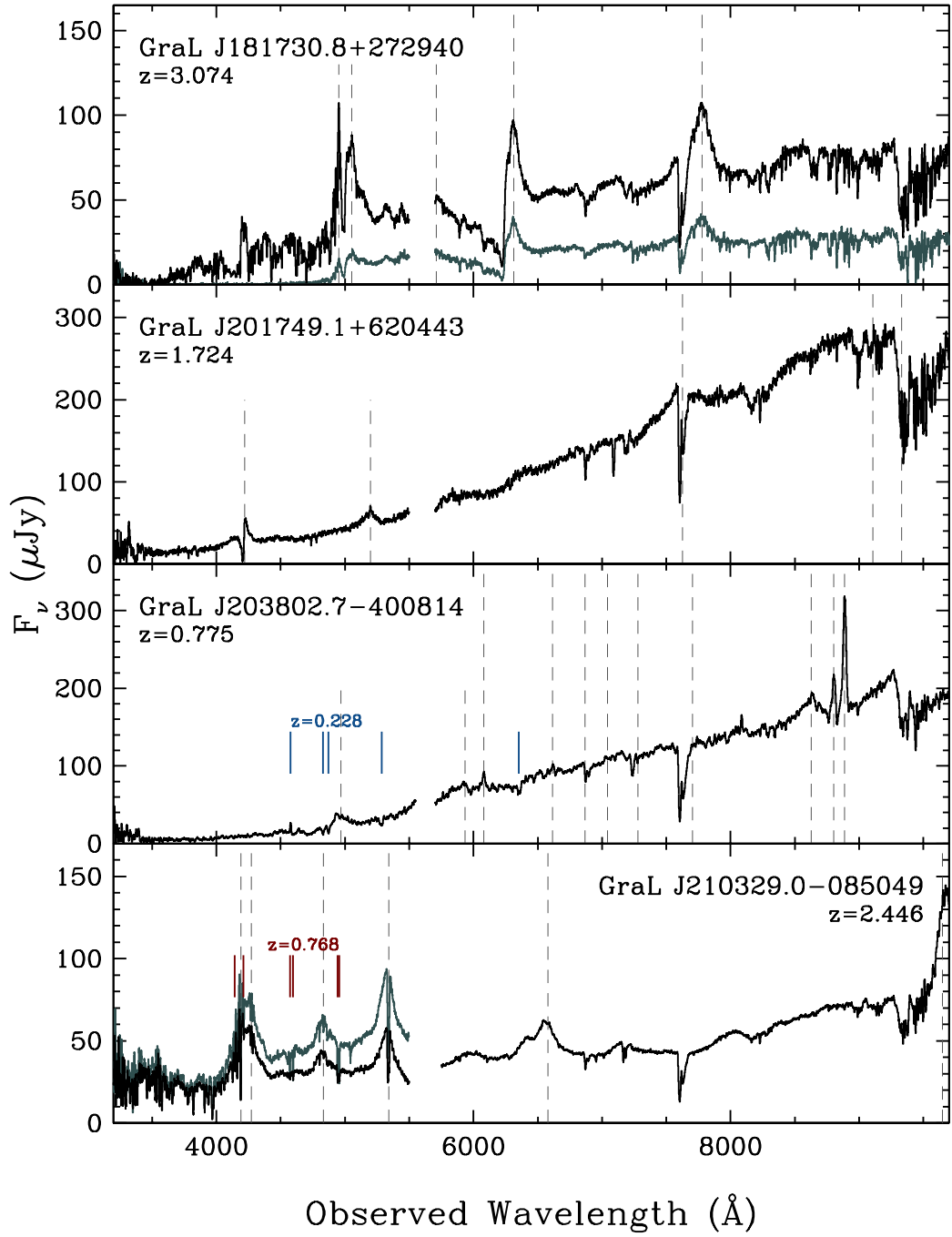}
\caption{Spectra of the final four confirmed quadruply-imaged GraL quasars, as per Figure~2.
\label{fig:spectra3}}
\end{center}
\end{figure*}

%
% FIGURE 5 - ASTERISM IMAGE STAMPS
\begin{figure*}
% \begin{figure*}[!ht]
\begin{center}
\includegraphics[width=1.0\textwidth]{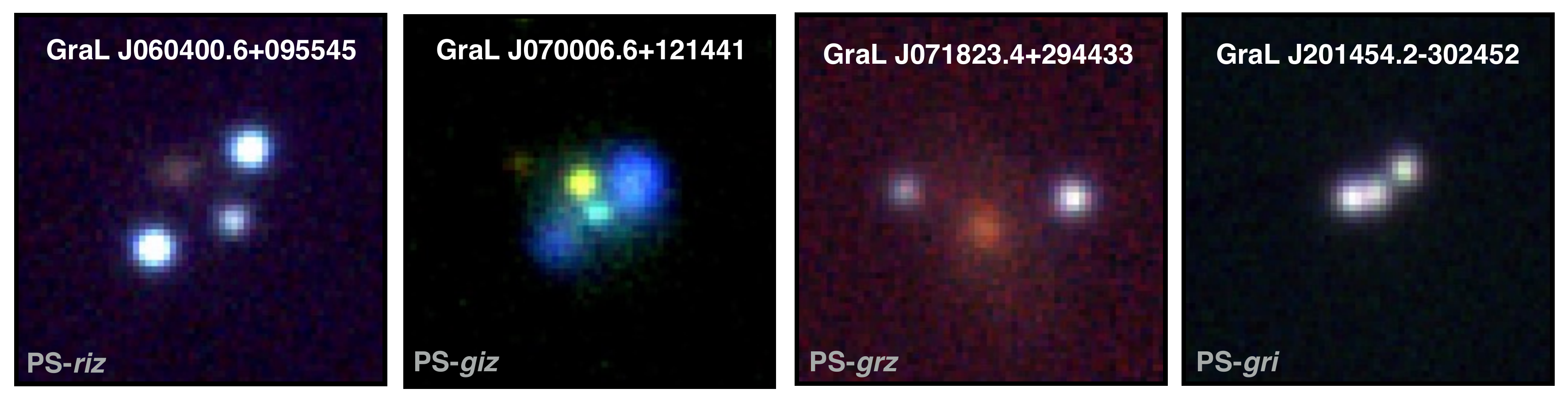} 
\caption{False-color images, 15\arcsec\ on a side, of example GraL lensed quasar candidates that spectroscopic follow-up disproved, or invalidated.  North is up, and east is to the left. Bottom left corner of each image indicates which PanSTARRS (PS) filters were mapped to the RGB colors.
\label{fig:images2}}
\end{center}
\end{figure*}

% FIGURE 6:  REJECT SPECTROSCOPY
\begin{figure*}[!htp]
\begin{center}
\includegraphics[width=1.8\columnwidth]{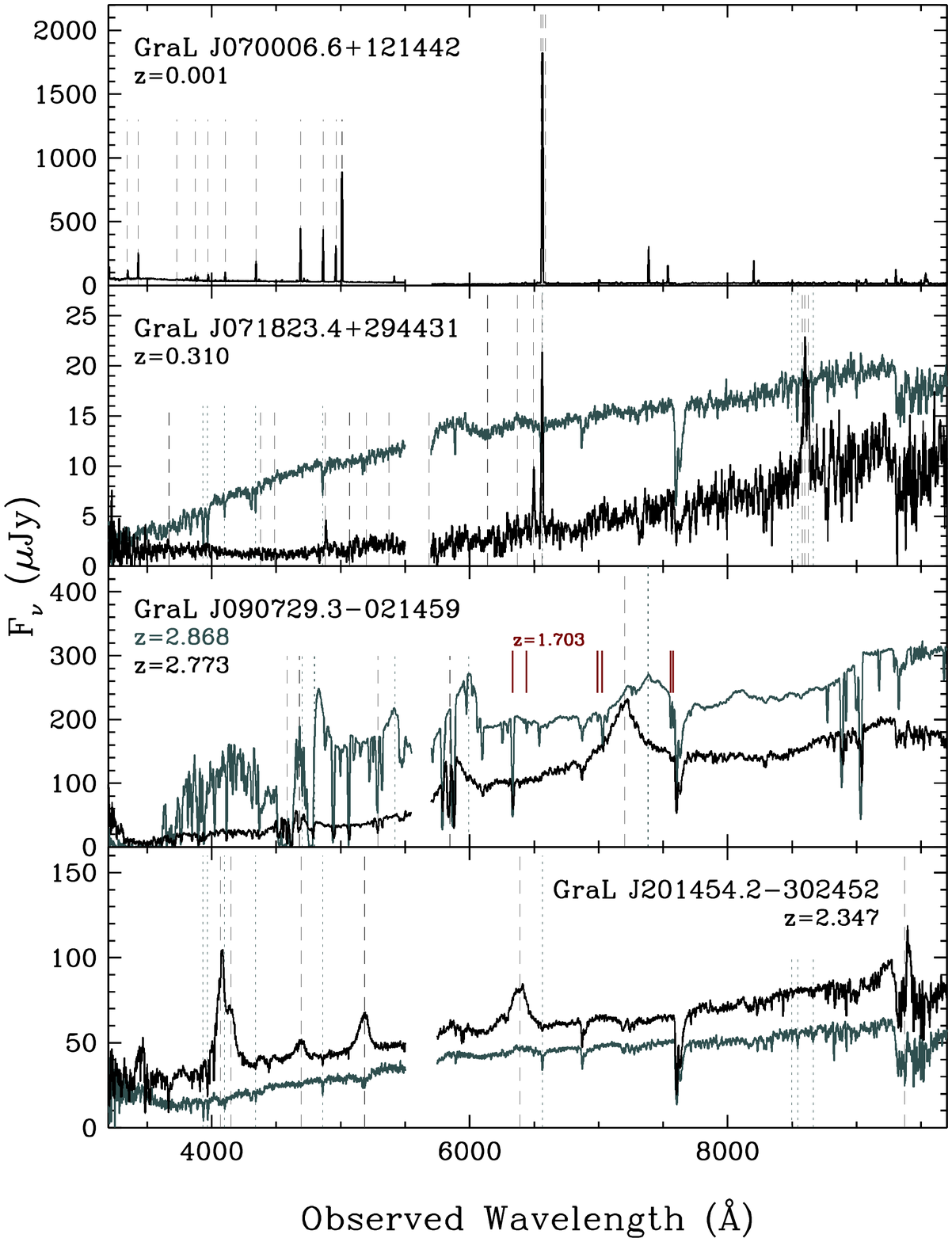}
\caption{Spectra of four GraL lensed quasar candidates that proved to be interlopers. Labels are as per Figure~2, with the addition of \ion{He}{2}~$\lambda 4686$, H$\alpha$, and [\ion{N}{2}]~$\lambda \lambda 6548, 6584$.  Dotted grey vertical lines show key spectral features for the Galactic sources in the second and fourth panels (in order, Ca~HK $\lambda\lambda 3933, 3967$, H$\delta$, H$\gamma$, H$\beta$, H$\alpha$, and the calcium triplet $\lambda \lambda \lambda 8448, 8498, 8542$).
\label{fig:spectra4}}
\end{center}
\end{figure*}

\section{Confirmed Asterisms}

In this section we discuss lensed quasar candidates that were disproved by spectroscopy.  Figures~5 and \ref{fig:spectra4} present examples of a few such sources, and Table~6 lists all of the invalidated GraL lens candidates (i.e., asterisms). We represent details on a few of the more interesting interlopers in our sample below. 

% \com{LD}{Like in the previous section, I think it worth to mention which technique was used for identifying these asterisms. FYI, the ERT were used for J0604+0955, J1226-4542 and J2014-3024. Also J0700+1214, J0718+2944, J0907-0214 were clearly doubly-imaged candidates, so should we discuss them in this quad paper? Potentially J155623264+300443480 can be interesting to discuss here (or add an `inconclusive' subsection).}

% [{\it Double check a few examples where mountain reduction suggested a lens, but my analysis didn't support that -- i.e., J2147+0719 (quasar + star), J1036$-$1335 (galaxy pair, one active), J1515$-$0322 (co-discovery w/ Rusu?), J1556+3004 (merging galaxies or lens?).}]

{\bf GraL~J060400.6+095545 ---}  This source, identified from an ERT analysis, looked like a promising large quadruply-imaged quasar, with the classic diamond configuration (see Figure~\ref{fig:images2}).  The \wise\ colors indicate the presence of an AGN.  However, the spectrum obtained at PA=135\deg\ reveals that the NW and SE sources are both Galactic.  Looking at the image suggests that the NE source, which looks softer and redder than the other components, is likely a galaxy --- and an active galaxy given the \wise\ colors --- while the other three components of this lens quasar candidate are instead Galactic.  Similar to GraL J071823.4+294431 discussed below, this source appears to correspond to our primary interloper:  the chance superposition of a distant AGN with foreground Galactic stars in a configuration that is suggestive of a gravitationally lensed quasar.

{\bf GraL~J070006.6+121442 ---} This source, identified as a candidate doubly-imaged quasar, appears as a resolved blue source, approximately 5\arcsec\ across, with a compact blue core and lobe-like structures stretching from the NW to the SE (see Figure~\ref{fig:images2}).  A redder point source is also evident to the NE.  This galaxy has very red mid-IR colors ($W1 - W2 = 1.42$), indicating a likely AGN, and the spectrum we obtained at PA=29\deg\ shows narrow, high equivalent emission lines from hydrogen, helium, nitrogen, and oxygen (see Figure~\ref{fig:spectra4}, top panel).  At first glance, the spectrum is suggestive of a low-redshift ($z=0.001$), low-metallicity starburst, similar to the low-metallicity blue dwarf galaxies reported, for example, by \citet{Griffith:11}, and a known contaminant to the \wise\ AGN selection \citep[e.g.,][]{Hainline:16}.  However, the \ion{He}{2}~$\lambda 4686$ and [\ion{Ne}{5}]~$\lambda 3426$ lines are unusually strong in this source, suggestive of an AGN.  \citet{Izotov:12} report on a sample of eight blue compact dwarf galaxies with [\ion{Ne}{5}] emission that they argue are likely associated with several hundred km~s$^{-1}$ radiative shocks.  However, in that sample, the \ion{He}{2} strengths are 1-4\%\ of the H$\beta$ strength and [\ion{Ne}{5}] is less than 1\%\ of the H$\beta$ strength.  In contrast, \ion{He}{2} is stronger than H$\beta$ in GraL~J070006.6+121442, while [\ion{Ne}{5}] is $\sim 90\%$ of the H$\beta$ strength.  This suggests that GraL~J070006.6+121442 might be a good candidate for hosting an active nucleus in a nearby dwarf galaxy \citep[e.g.,][]{Reines:13}.

{\bf GraL J071823.4+294431 ---}  This candidate doubly-imaged quasar, shown in Fig.~\ref{fig:images2}, appears as an enticing configuration consisting of a red galaxy flanked by two point sources, suggestive of an elliptical galaxy lensing a background quasar.  Spectroscopy instead reveals the flanking sources to be Galactic stars while the galaxy is an obscured AGN with a very high [\ion{O}{3}]:H$\beta$ ratio and relatively narrow H$\alpha$ (Figure~\ref{fig:spectra4}, second panel; Galactic source has been scaled down in flux by 66\%). This is the most aesthetic example of one of our most common failure modes:  an AGN closely paired with Galactic stars.  The AGN is responsible for the red mid-IR colors leading to the candidate selection, while asterisms with Galactic point sources lead to the \gaia\ selection of a lens candidate.

{\bf GraL~J090729.3$-$021449 ---}  This candidate doubly-imaged quasar turns out to consist of two distinct quasars at $z = 2.773$ and $z = 2.868$, separated by $\approx 2\arcsec$ (Figure~\ref{fig:spectra4}, third panel).  Absorption from the foreground quasar is evident in the background quasar, providing an opportunity to study the environment and feedback from the foreground AGN \citep[e.g.,][]{Hennawi:06}.  In addition, there are  absorption systems foreground to both quasars, such as the $z = 1.703$ systems shown in Figure~\ref{fig:spectra4}, providing an opportunity to study the spatial extent and kinematics of quasar absorption line systems.

{\bf GraL~J122628.5$-$454209 ---} This was one of the three new
candidate quadruply-imaged quasars presented in Krone-Martins \etal\, (2018),
along with the now confirmed GraL~J113100.0$-$442000 and
GraL~J203802.7$-$400814 systems discussed in \S~4.  With $P_{\rm ERT} = 0.22$,
this source had the lowest ERT probability of all the
sources discussed in that paper, which also included several known,
confirmed quadruply-imaged quasars; these confirmed lens systems
all had $P_{\rm ERT} \simgt 0.95$.  The low $P_{\rm ERT}$ was thought
due a combination of the optical faintness of several of the lensed
images, as well as the proximity of this system to a bright ($G =
16.3$) star at a distance of only $\sim 2\arcsec$.  Observations
obtained during the first spectroscopic observing run for this
project identified the Galactic nature of GraL~J122628.5$-$454209.
This is consistent with the blue mid-IR colors of the source, $W1-W2
= 0.2$, which is unresolved by {\it WISE}. In contrast, the two
confirmed lenses from that paper have red, $W1 - W2 > 1$ colors,
indicative of an AGN \citep[e.g.,][]{Stern:12}.

{\bf GraL~J201454.2$-$302452 ---}  This source, shown in Fig.~\ref{fig:images2}, was presented in \cite{Delchambre:19} as a candidate lensed quasar with three \gaia-detected lensed images, an ERT probability of $P_{\rm ERT} = 0.88$, and a compact maximum separation of  2\farcs5.
Spectroscopic follow-up shows the two eastern sources to be a doubly-imaged quasar at $z = 2.347$, while the western source is a Galactic star.  Figure~\ref{fig:spectra4} (bottom panel) shows one of the quasar spectra and the western Galactic source spectrum. Due to imperfect source deblending, faint features from the Galactic star are imprinted on the quasar spectrum, and vice versa.

%{\bf GraL~222320.3+260549 ---} This candidate is another example of the common failure mode of a spectroscopically confirmed quasar in close proximity to a Galactic star (or two stars, in this case).  What makes this case notable is that the quasar is at $z = 3.840$, the most distant source in this paper, with an $i$-band magnitude of $i = 18.86$.  

\section{Lens modelling}

We have performed gravitational lens modeling of each of the confirmed quad lensed systems using  two standard families of mass distribution: a singular isothermal sphere (SIS) and a singular isothermal ellipsoid \citep[SIE;][]{Kormann:94}. The parameters of these models are the Einstein radius $\theta_E$, lens galaxy position, and, in the case of the SIE, the lens ellipticity $e=1-q$ (where $q$ is the axis ratio) and position angle PA$_e$. In both cases, an additional shear term has been added to the model to account for the statistical gravitational impact of galaxies along the line-of-sight towards the lens system. For the SIS lens model, this shear term also accounts for ellipticity in the lens.  The shear term parameters are its amplitude $\gamma$ and its position angle PA$_g$ (reported E of N, and pointing towards the mass producing the shear). The modeling has been performed using the public modeling software \texttt{lensmodel} \citep{Keeton:01}. The following observables were used to constrain the lens models: the relative astrometry between the lensed images, and the observed flux ratios. For each system, the \gaia\ astrometry and photometry were used except for GraL~J065904.1$+$162909, for which positions and fluxes come from SDSS. We used the formal uncertainty on positions but enforced a minimum astrometric uncertainty of 0\farcs002 to account for possible perturbation of image positions caused by structures in the lens potential not described by our model (e.g., dark matter subhaloes of $M \leq 10^{10} M_\odot$). We also enlarged the uncertainty on the flux ratio to account for deviations from any macro model as caused by microlensing, millilensing, intrinsic variability, and/or differential dust reddening. Table~\ref{table:data_for_models} in the Appendix provides a summary of the data used. 

The positions of all four lensed images were available for half of the sample. For those systems, we enlarged the flux ratio relative uncertainty to 20\%. For the other five systems, data were available for only a triplet of images (see Tables~\ref{table:SIS} and \ref{table:SIE}), and a flux ratio relative uncertainty of 10\% has been included. For triplets, we have also guessed the position of the fourth image to initiate the modeling. Finally, for a few systems, we have complemented this information with measurements from ancillary data. We have used the PanSTARRS position of the lensing galaxy for GraL J165105.3$-$041725, and of the fourth lensed image for GraL 065904.1+162909.   

\subsection{Results}

\begin{figure*}[!ht]
\begin{center}
\includegraphics[width=1.0\textwidth]{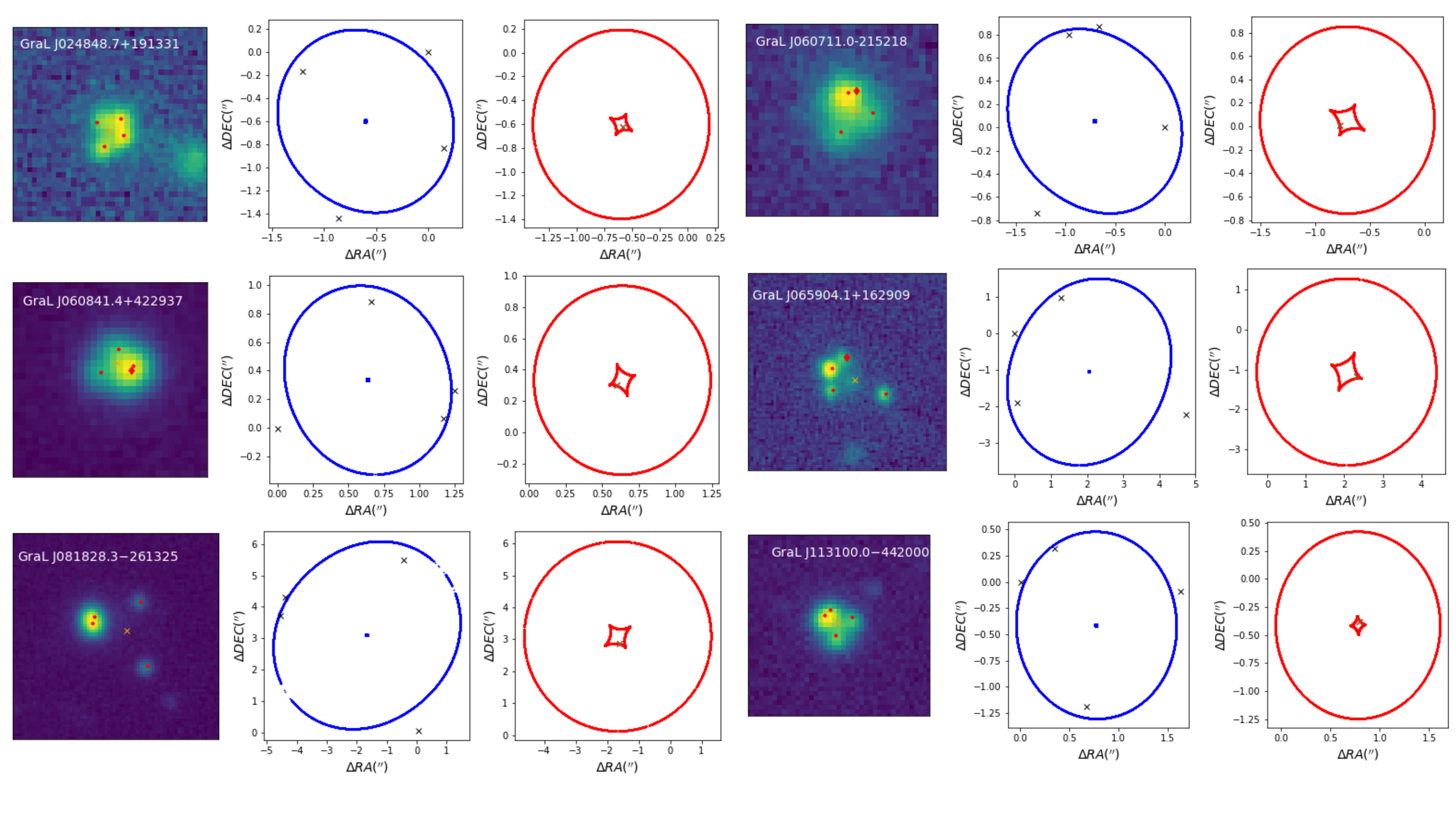} 
\caption{Summary of the lens models (SIS+$\gamma$) for the first six confirmed quadruply-imaged GraL quasars: the left panel shows the $i$-band image, with \gaia\ point-source positions indicated with a red dot and a red diamond when the position is predicted by the model. The middle panel shows the critical curves with image positions indicated by crosses, and the right panel the caustics and the predicted position of the source indicated by a cross. 
\label{fig:models_1}}
\end{center}
\end{figure*}

\begin{figure*}[!ht]
\begin{center}
\includegraphics[width=1.0\textwidth]{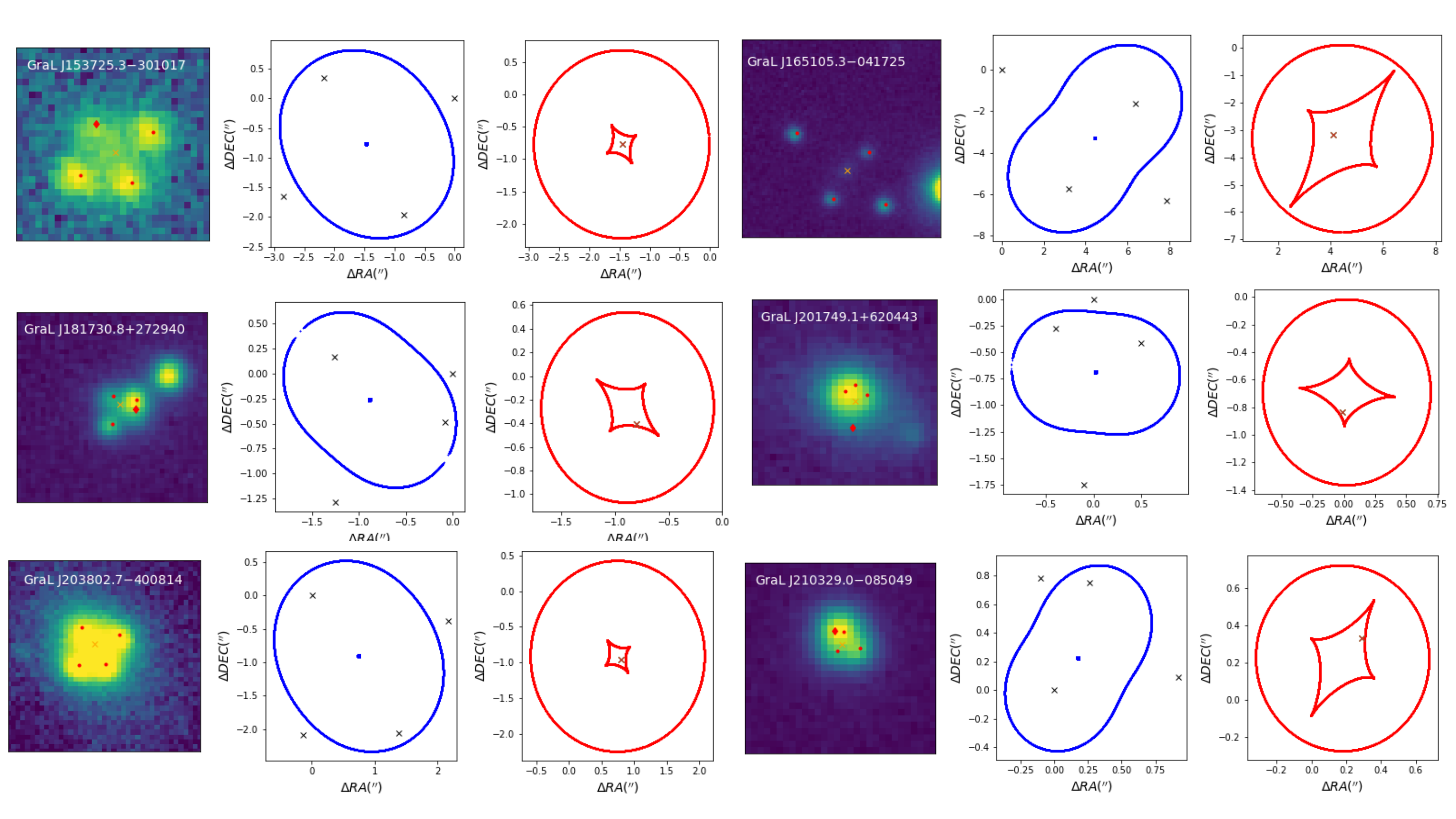} 
\caption{Summary of the lens models (SIS+$\gamma$) for the last six confirmed quadruply-imaged GraL quasars, as per Figure~7.
\label{fig:models_2}}
\end{center}
\end{figure*}

The results of the modeling are presented in Tables~\ref{table:SIS} and \ref{table:SIE}, and Figures~\ref{fig:models_1} and \ref{fig:models_2}. Models for each systems are discussed below. The most robust parameters from this kind of lens modeling are the Einstein radius $\theta_E$ and the position angle of the lensing galaxy PA$_e$ when four lensed images are observed \citep[e.g.,][]{Sluse:12}. Models that involve only three lensed images are more uncertain. Because SIE$+\gamma$ models are under-constrained when only three lensed images are observed, we chose not to present them. 

%In particular, the SIE$+\gamma$ models are under-constrained in the case of a triplet (i.e., more parameters than observable), and inferred parameters must be considered with the understanding that the modeling solution is not unique. We briefly comment below on the individual lens models. 

{\bf GraL J024848.7+191331 ---} A model based on {\it HST} imaging of this system was published in \cite{Shajib:19}. The {\it HST} data confirm the image configuration and Einstein radius of our simple model based on only three lensed images detected by \gaia. %\comDS{JS Since the relative positions of the 4th lensed image are known, why not to use them and indicate the appropriate reference (cf. HST, PS)? JS } DS: GRAL meeting Sept. 2020 => Only GAIA astrometry 

{\bf GraL J060710.8$-$215217 ---} Our model for this compact quad consists of a fold-configuration lens. Our SIS+$\gamma$ model requires a moderate shear ($\gamma \sim 0.08$), but this shear could  effectively be caused by the lens ellipticity. This model implies that the brightest component visible in the ground-based imaging is in fact composed of a merging pair, with the two, roughly equal brightness components unresolved by \gaia. 

{\bf GraL J060841.4+422937 ---} The image configuration and model obtained for this system are similar to GraL J060710.8$-$215217. The fourth image is predicted to be $\sim 0\farcs2$ from the brightest component detected in \gaia\ data. 

{\bf GraL J065904.1+162909 ---} This large separation system ($\theta_E \sim 2\farcs4$) is not easily reproduced with a SIS+$\gamma$ model. A substantial shear and a very large ellipticity are required to reasonably reproduce the image positions. However, the flux ratios are poorly reproduced by this model. Since the flux ratios between the images forming the bright triplet substantially deviates from the asymptotic relation linking ``cusp'' images, we suspect that substantial microlensing is at work in (at least) one of those images \citep[e.g.,][]{Mao:98}. On the other hand, the extreme value found for the ellipticity and large Einstein radius ($\theta_E = 2\farcs37$) are indicative of a missing ingredient in the mass model, either a nearby massive galaxy, or a galaxy group/cluster along the line-of-sight.   

{\bf GraL J081828.3$-$261325 ---} This is the system with the second largest Einstein radius ($\theta_E = 3\farcs01$). Both shear and lens ellipticity are required to reproduce the image configuration, as shown by the large $\chi^2$ associated with the SIS$+\gamma$ model. That model is unable to reproduce the observed image positions. This indicates that a galaxy group or cluster may lie along the line-of-sight towards this system. 

{\bf GraL J113100.0$-$442000 ---} This system was  modeled in GraL~IV \citep{Wertz:19}. The models presented here agree with the previous ones. 

{\bf GraL J153725.3$-$301017 ---} This system has a rather large $\theta_E \sim 1\farcs5$. While the model is very uncertain since it is based on three lensed images only, it predicts positions of the fourth image and of the lensing galaxy that are qualitatively compatible with the PanSTARRS imaging. An SIE model (without shear) was presented in \cite{Lemon:19}. They find $\theta_E = 1\farcs48$, in agreement with our SIS$+\gamma$ model. %\comDS{JS Here also, since the relative positions of the 4th lensed image are known, why not to use them and indicate the appropriate reference (cf. HST, PS)? JS} DS: GRAL meeting Sept. 2020 => Only GAIA astrometry 

{\bf GraL 165105.3$-$041725 ---} The lensed images of this very large separation system are well reproduced by our simple models provided that a large external shear is included, as one already expects from the elongated cross-like morphology. In addition, the centroid of the mass model is found to be significantly offset (0\farcs15) from the observed position of the main lensing galaxy. Note that this is the only system for which we have used a lens galaxy position for the modeling. These properties indicate the presence on the line-of-sight towards this system of one or several massive perturbers, most probably a galaxy group or cluster, possibly at the same redshift as the main deflector. This situation is typical of very large separation systems \citep[e.g.,][]{Walsh:79, Inada:03, Rusu:13}. 

{\bf GraL J181730.8+272940 ---} The best SIS$+\gamma$ model that reproduces the triplet of \gaia-detected images is a fold-configuration model for which the brightest component is a merging of two images. However, based on multi-band and high-resolution images, \cite{Rusu:18} showed that the fourth lensed image suffers from strong extinction by the lensing galaxy, such that this system has effectively a cross-like configuration. This configuration cannot be reproduced with a SIS+$\gamma$ model, but requires a more complex mass distribution and external shear, as shown by \cite{Rusu:18}. The lens models presented by those authors supersede the ones presented here. %\comDS{Do we instead make a model with the fourth image guessed at the right position ? JS Yes! DS ; In fact it does not work (and I did it but forgot) }

{\bf GraL J201749.1+620443 ---} Our lens model consists of a short-axis cusp configuration, with a fourth lensed image predicted opposite to the observed triplet. Despite the small Einstein radius ($\theta_E = 0\farcs69$), a very large shear is needed to accommodate this model ($\gamma = 0.22$). The latter could be associated with the nearby red object observable south-west of the lens, which is roughly along the direction producing the shear in our model. % \comDS{JS Model parameters are missing in Tables 3 and 4 JS } DS: Fixed

{\bf GraL J203802.7$-$400814 ---} This system is well reproduced by our simple models. The SIE$+ \gamma$ model indicates a moderate shear and ellipticity. The presented models are in qualitative agreement with previously published ones \citep{Agnello:18, Shajib:19}. The model presented in \cite{Shajib:19}, constrained by {\it HST} imaging that detects the host galaxies, supersedes the models presented here. 
%\comDS{JS Here also, since additional observing constraints exist, I would be in favour of incorporating all of these in the modeling (cf. HST, DES)? JS } DS: GRAL meeting Sept. 2020 => Only GAIA astrometry
%At the exception of the Einstein radius\footnote{We derive $\theta_E = 1\farcs36$ while \cite[][]{Agnello:18} reports $\theta_E = 1\farcs26$)}, the reported model parameters are compatible with model (a) in Table 3 of \cite[][]{Agnello:18}, which was based on DES imaging.  

{\bf GraL J210329.0$-$085049  ---} Our model consists of a fold-configuration lens for which the brightest component visible in ground-based imaging is in fact composed of two merged sources. This model requires a large extrinsic shear ($\gamma = 0.26$) and predicts a fourth lensed image as bright as the observed one. The absence of this fourth image in the \gaia\, data suggests that this model and the predicted image morphology is erroneous. 

% FIGURE 7 - 
% Lens models summary `GRAL_model_new_lenses_May_2020.ipynb' + ppt to make composite

% FIGURE 8

%
% TABLE 3:  SIS MODELING
% Table generated in `GRAL_model_new_lenses_May_2020.ipynb' w. models from lensmodel + ppt to make composite
% \scriptsize
\begin{deluxetable*}{lcccccc}
% \tablewidth{0pt}
\tablecaption{Results of the SIS + $\gamma$ model.}
\tablehead{
\colhead{Name} &
\colhead{$\theta_E$} &
\colhead{$\gamma$} &
\colhead{PA$_g$} &
\colhead{$\chi^2_{\rm ima}$} &
\colhead{$\chi^2_{\rm flux}$} &
\colhead{$\chi^2_{\rm gal}$} \\
\colhead{} &
\colhead{(\arcsec)} &
\colhead{} &
\colhead{($\deg$)} &
\colhead{} &
\colhead{} &
\colhead{}}
\startdata
GraL J024848.7+191331${\dagger}$    & 0.81 & 0.06 &  61 &     0.06 &  0.28 & \nodata \\
GraL J060711.0-215218${\dagger}$ & 0.82 & 0.10 & 61 & 0.89 & 0.46 & \nodata \\
GraL J060841.4+422937${\dagger}$ & 0.62 & 0.08 & 20 & 1.09 & 3.43 & \nodata \\
GraL J065904.1+162909    & 2.37 & 0.10 & -24 &   118.13 & 69.19 & \nodata \\
GraL J081828.3$-$261325 & 3.01 & 0.08 & -53 & 12697.06 &  5.66 & \nodata \\
GraL J113100.0$-$442000 & 0.85 & 0.05 &   4 &    86.18 &  4.73 & \nodata \\
GraL J153725.3$-$301017${\dagger}$ & 1.48 & 0.10 &  30 &     0.02 &  0.04 & \nodata \\
GraL J165105.3$-$041725${\dagger\dagger}$& 3.46 & 0.31 & -38 &    32.05 &  7.11 & 493.06 \\
GraL J181730.8+272940${\dagger}$   & 0.82 & 0.19 &  51 &     0.81 &  5.55 & \nodata \\
GraL J201749.1+620443${\dagger}$   & 0.69 & 0.22 &  85 &     1.65 &  0.17 & \nodata \\
GraL J203802.7$-$400814 & 1.36 & 0.09 &  35 &    21.08 &  1.51 & \nodata \\
GraL J210329.0$-$085049${\dagger}$ & 0.51 & 0.26 & -30 &     1.69 &  0.18 & \nodata
\enddata
\label{table:SIS}
\tablecomments{${\dagger}$: Models based only on three lensed image positions. ${\dagger\dagger}$: Model includes lens galaxy position from PanSTARRS.} 
% \com{LD}{If the galaxy position is not used it is clearer to set $\chi^2_{\rm gal}$ as an empty field instead of setting it to zero.} }
\end{deluxetable*}

% TABLE 4:  SIE MODELING
% Table generated in `GRAL_model_new_lenses_Oct_2019.ipynb' w. models from lensmodel
% \scriptsize
\begin{deluxetable*}{lcccccccc}
% \tablewidth{0pt}
\tablecaption{Results of the SIE + $\gamma$ model.}
\tablehead{
\colhead{Name} &
\colhead{$\theta_E$} &
\colhead{$e$} &
\colhead{PA$_e$} &
\colhead{$\gamma$} &
\colhead{PA$_g$} &
\colhead{$\chi^2_{\rm ima}$} &
\colhead{$\chi^2_{\rm flux}$} &
\colhead{$\chi^2_{\rm gal}$} \\
\colhead{} &
\colhead{(\arcsec)} &
\colhead{} &
\colhead{($\deg$)} &
\colhead{} &
\colhead{($\deg$)} &
\colhead{} & 
\colhead{} &
\colhead{} }
\startdata
%GraL J024848.7+191331   & 0.76 & 0.37 &  -1 & 0.19 &  81 & 0.00 &  0.00 & 0.00 \\
GraL J065904.1+162909   & 2.43 & 0.70 & -18 & 0.10 &  83 & 4.10 & 33.34 & \nodata \\
GraL J081828.3$-$261325 & 2.87 & 0.36 &  10 & 0.28 & -61 & 0.01 &  8.22 & \nodata \\
GraL J113100.0$-$442000 & 0.85 & 0.10 &  25 & 0.04 & -13 & 0.01 &  4.29 & \nodata \\
%GraL J153725.3$-$301017 & 1.46 & 0.10 & -73 & 0.13 &  27 & 0.03 &  0.00 & 0.00 \\
GraL J165105.3$-$041725 & 3.56 & 0.39 & -38 & 0.24 & -38 & 38.68 & 12.21 & 379.52 \\
%GraL J181730.8+272940   & 0.84 & 0.15 &  79 & 0.14 &  44 & 0.64 & 0.00 & 0.00 \\
%GraL J201749.1+620443   & 0.73 & 0.18 & -49 & 0.24 &  76 & 1.83 & 0.07 & 0.00 \\
GraL J203802.7$-$400814 & 1.35 & 0.21 & -58 & 0.13 &  33 & 0.00 & 0.35 & \nodata 
%GraL J210329.0$-$085049 & 0.51 & 0.24 & -25 & 0.20 & -33 & 1.04 & 0.01 & 0.00
\enddata
\tablecomments{Models for which four lensed image positions are detected.}
\label{table:SIE}
\end{deluxetable*}

\section{Conclusions}

We present the results of spectroscopic efforts to verify candidate strongly lensed quasars identified using machine learning methods on \gaia\ astrometric and photometric data, supplemented with mid-IR photometry from \wise\ and optical light curves from CRTS and Palomar observatory.  We present 12 confirmed quadruply-imaged quasars, seven of which are first reported here.  All of the confirmed lenses were either first reported as candidates by the GraL collaboration or were independently identified by GraL and other collaborations. We also present one doubly-imaged quasar, GraL~J201454.2$-$302452, that was initially selected as three-image lens. A companion paper, \citet{KroneMartins:20}, reports on the selection and confirmation of doubly-imaged quasars from these efforts.  As an aid to future investigations, we also present the spectroscopic failures, where follow-up observations generally revealed the candidates to be either purely Galactic or the chance asterism of Galactic stars and a background, typically active, galaxy.

\citet{Delchambre:19}, a previous paper from our collaboration, presented 15 candidate quadruply-imaged quasars.  What do our spectroscopic observations reveal about the success rate for these candidates?  Of the 15 candidates, we now have spectroscopic follow-up of 12; the three remaining candidates are in the southern sky and are challenging to inaccessible from our primarily northern follow-up facilities at Maunakea and Palomar Mountain (one candidate is at $-37\deg$ declination, and the other two have declinations below $-50\deg$).  Of the 12 observed sources, six are confirmed as quadruply-imaged quasars and a seventh source, GraL~J201454.2$-$302452, is a doubly-imaged quasar with a close Galactic interloper that masqueraded as a third lensed image (see \S~5).  Two more candidates are asterims of a quasar and Galactic stars, and the final three candidates (which include both the highest and the lowest ERT probability candidates from that paper) are purely Galactic.  Notably, all three Galactic systems have blue mid-IR colors ($W1 - W2 < 0$), while all nine confirmed quasars have red mid-IR colors, $W1 - W2 \simgt 0.6$, though two are not sufficiently red to have been identified using the \citet{Stern:12} mid-IR color criterion (i.e., $W1 - W2 \geq 0.8$).  Of the three remaining candidates from \citet{Delchambre:19} lacking spectroscopic follow-up, only GraL~J053036.9$-$373011 has \wise\ colors indicative of a quasar.  With $P_{\rm ERT} = 0.98$ and a very compact configuration with three \gaia\ sources within 1\farcs0, this is the highest priority source for future spectroscopy.  Were we to impose a modest and simple mid-IR color criterion of $W1 - W2 > 0.5$ on the \gaia\ lensed quasar candidates from \citet{Delchambre:19}, then our spectroscopy would confirm that all 9 observed candidates (100\%) include an active galaxy, with seven (78\%) including a lensed quasar and six (67\%) confirmed as the quadruply-imaged quasars sought by \citet{Delchambre:19}.

This paper presents the twelve confirmed GraL quadruply-imaged quasars as of Summer 2020, which accounts for approximately a 20\%\ increase in the number of confirmed quadruply-imaged quasars.  This work sets the stage for multi-wavelength follow-up activities. Besides photometric monitoring which will enable using these sources as cosmological probes, we also have ongoing follow-up campaigns that include high-resolution radio follow-up, X-ray observations, and near-infrared adaptive optics integral field unit spectroscopy to improve modeling and enable scientific exploitation of these rare systems.  The results presented here have also guided revised machine learning approaches by the GraL team to identify new lensed quasars based on the forthcoming \gaia\ data releases and public data sets, such as the recently published \gaia\ Early Data Release 3 \citep{Gaia:20}. The hunt continues.

% Delchambre+19 candidates:
%[4] 214110146+314107480 4 -- stars; w1-w2 < 0
%[8] 053036992–373011003 3 -- no spectroscopy; w1-w2 > 1
%[11] 153725327–301017053 3 -- w1-w2 = 0.698
%[12] 113100013–441959935 4 -- w1-w2 > 1
%[15] 081602164–530722970 4 -- no spectroscopy; w1-w2 = 0.07
%[16] 175443398+214054818 3 -- w1-w2 = 0.880
%[17] 065904044+162908685 3 -- w1-w2 = 0.595
%[18] 182244519–541451730 4 -- no spectroscopy; w1-w2 < 0
%[19] 054934271+051814610 3 -- w1-w2 = 0.998
%[20] 075933618–173212537 3 -- stars; w1-w2 < 0
%[23] 181730853+272940139 3 -- w1-w2 = 0.834
%[25] 024848742+191330571 3 -- w1-w2 > 1
%[26] 201454150–302452196 3 -- w1-w2 = 0.985
%[28] 201749047+620443509 3 -- w1-w2 > 1
%[30] 011559515+562506671 3 -- stars; w1-w2 < 0

\acknowledgements

% \com{LD}{The title of the acknowledgement section does not appear here...}
% \comDS{DSl: We have two DS in the team ! Daniel and myself. How do you wish to distinguish ? } 

The majority of this manuscript preparation took place during the COVID-19 global pandemic. The authors would like to thank all those who risked their lives as essential workers in order for us to safely continue our work from home.  We also thank the staff at the various observatories who assisted in the acquisition of the data presented herein. We gratefully acknowledge Amy Reines for useful discussions about GraL~J070006.6+121442, which is suggestive of an AGN in a low-$z$, low-$Z$ galaxy.  The work of D.S. was carried out at the Jet Propulsion Laboratory, California Institute of Technology, under a contract with NASA. S.G.D., A.D., M.J.G., and A.M. are supported by the NSF grants AST-1518308 and AST-1815034, and the NASA grant 16-ADAP16-0232. AKM acknowledge the support from the Portuguese Funda\c c\~ao para a Ci\^encia e a Tecnologia (FCT) through the Portuguese Strategic Programme UID/FIS/00099/2019 for CENTRA and through grants SFRH/BPD/74697/2010 \& PTDC/FIS-AST/31546/2017. This project has received funding from the European Research Council (ERC) under the European Union’s Horizon 2020 research and innovation programme (grant agreement No 787886).

Some of the data presented herein were obtained at the W. M. Keck Observatory, which is operated as a scientific partnership among the California Institute of Technology, the University of California and the National Aeronautics and Space Administration. The Observatory was made possible by the generous financial support of the W. M. Keck Foundation.  The authors wish to recognize and acknowledge the very significant cultural role and reverence that the summit of Maunakea has always had within the indigenous Hawaiian community.  We are most fortunate to have the opportunity to conduct observations from this mountain.

Based on observations obtained at the Hale Telescope, Palomar Observatory as part of a continuing collaboration between the California Institute of Technology, NASA/JPL, Yale University, and the National Astronomical Observatories of China.

Some of the data presented herein were obtained at the international Gemini Observatory, a program of NSF’s NOIRLab, which is managed by the Association of Universities for Research in Astronomy (AURA) under a cooperative agreement with the National Science Foundation. on behalf of the Gemini Observatory partnership: the National Science Foundation (United States), National Research Council (Canada), Agencia Nacional de Investigaci\'{o}n y Desarrollo (Chile), Ministerio de Ciencia, Tecnolog\'{i}a e Innovaci\'{o}n (Argentina), Minist\'{e}rio da Ci\^{e}ncia, Tecnologia, Inova\c{c}\~{o}es e Comunica\c{c}\~{o}es (Brazil), and Korea Astronomy and Space Science Institute (Republic of Korea).

This work has made use of data from the ESA mission \gaia, processed by the \gaia\ Data Processing and Analysis Consortium (DPAC). Funding for DPAC has been provided by national institutions, in particular the institutions participating in the \gaia\ Multilateral Agreement. This publication makes use of data products from the {\it Wide-field Infrared Survey Explorer}, which is a joint project of the University of California, Los Angeles, and the Jet Propulsion Laboratory/California Institute of Technology, funded by the National Aeronautics and Space Administration. This research has made use of the Aladin Sky Atlas and the SIMBAD database, both developed and operated at Centre de données astronomiques de Strasbourg (CDS) at Strasbourg Observatory, France.  This research has made use of the NASA/IPAC Extragalactic Database (NED), which is operated by the Jet Propulsion Laboratory, California Institute of Technology, under contract with the National Aeronautics and Space Administration.

% \section{References}
\bibliographystyle{apj.bst}
% \bibliography{bibliography}

\smallskip
{\it Facilities:} \facility{CRTS}, \facility{Gaia}, \facility{Gemini (GMOS)}, \facility{Keck:I (LRIS)}, \facility{NTT (EFOSC2)}, \facility{Palomar (DBSP)}, \facility{WISE}, \facility{ZTF}.

\smallskip
\copyright 2020.  All rights reserved.

\eject
\appendix

Table~5 presents the astrometric and photometric data used for the lens modeling described in \S~6; these data come from \gaia\ for all sources other than the western component of GraL~J065904.1+162909, and the listed \gaia\ photometry corresponds to the $G$-band mean magnitudes (Vega); for that one exception, the data comes from SDSS. Table~6 presents the GraL lens candidates which spectroscopic observations failed to confirm as lensed quasars.

% TABLE 5
% Table generated in `GRAL_model_new_lenses_May_2020.ipynb' w. input for lensmodel
% \scriptsize
\begin{deluxetable*}{cccccc}
% \tablewidth{0pt}
\tablecaption{Input data used for the modeling}
\tablehead{
\colhead{Name} &
\colhead{$\Delta$RA} &
\colhead{err $\Delta$RA} &
\colhead{$\Delta$Dec} &
\colhead{err $\Delta$Dec} &
\colhead{$G$} \\
\colhead{} &
\colhead{(\arcsec)} &
\colhead{(\arcsec)} &
\colhead{(\arcsec)} &
\colhead{(\arcsec)} &
\colhead{(mag)}}
\startdata
%\multicolumn{5}{l}{\textcolor{red}{LD: I would personally not repeat the lens identifier in the first column}} \\
GraL J024848.7+191331 & 0.0000 & 0.002 & 0.0000 & 0.002 & 20.41 \\
                      & 0.8553 & 0.004 & -1.4426 & 0.003 & 20.71 \\
                      & -0.1478 & 0.002 & -0.8383 & 0.002 & 20.41 \\
GraL J060711.0$-$215218 & 0.0000 & 0.002 & 0.0000 & 0.005 & 20.90 \\
                        & 1.2907 & 0.002 & -0.7385 & 0.003 & 20.86 \\
                        & 0.9647 & 0.002 & 0.7982 & 0.002 & 19.32 \\
GraL J060841.4+422937 & 0.0000 & 0.008 & 0.0000 & 0.010 & 20.13 \\
                      & -1.2462 & 0.002 & 0.2580 & 0.002 & 18.23 \\
                      & -0.6614 & 0.003 & 0.8846 & 0.003 & 19.86 \\
GraL J065904.1+162909 & 0.0000 & 0.002 & 0.0000 & 0.002 & 18.59 \\
                      & -4.7491 & 0.002 & -2.2357 & 0.002 & 20.05 \\
                      & -0.0849 & 0.002 & -1.9017 & 0.002 & 19.94 \\
                      & 1.1879 & 0.003 & 0.9889 & 0.010 & 20.00 \\
GraL J081828.3$-$261325 & 0.0000 & 0.002 & 0.0000 & 0.002 & 19.74 \\
                      & 4.3682 & 0.002 & 3.6783 & 0.002 & 17.58 \\
                      & 0.4577 & 0.002 & 5.5737 & 0.002 & 19.94 \\
                      & 4.4929 & 0.002 & 4.3175 & 0.002 & 17.52 \\
GraL J113100.0$-$442000 & 0.0000 & 0.002 & 0.0000 & 0.002 & 19.38 \\
                      & -1.6280 & 0.002 & -0.1000 & 0.002 & 20.13 \\
                      & -0.6888 & 0.002 & -1.1864 & 0.002 & 20.31 \\
                      & -0.3454 & 0.002 & 0.3246 & 0.002 & 19.32 \\
GraL J153725.3$-$301017 & 0.0000 & 0.002 & 0.0000 & 0.002 & 20.32 \\
 & 0.8423 & 0.002 & -1.9630 & 0.003 & 20.45 \\
 & 2.8427 & 0.002 & -1.6485 & 0.002 & 20.22 \\
GraL J165105.3$-$041725 & 0.0000 & 0.002 & 0.0000 & 0.002 & 19.60 \\
 & -3.2119 & 0.002 & -5.7522 & 0.002 & 19.49 \\
 & -7.8631 & 0.002 & -6.3014 & 0.002 & 18.99 \\
 & -6.3679 & 0.002 & -1.6598 & 0.002 & 20.04 \\
GraL J165105.3$-$041725-G & -4.5961 & 0.010 & -3.1504 & 0.010 & \nodata \\
GraL J181730.8+272940 & 0.0000 & 0.002 & 0.0000 & 0.002 & 18.93 \\
 & 1.2633 & 0.002 & 0.1584 & 0.002 & 20.72 \\
 & 1.2565 & 0.002 & -1.2832 & 0.002 & 20.07 \\
GraL J201749.1+620443 & 0.0000 & 0.002 & 0.0000 & 0.002 & 19.15 \\
 & -0.5011 & 0.020 & -0.4097 & 0.060 & 20.14 \\
 & 0.4052 & 0.009 & -0.2760 & 0.041 & 19.71 \\
GraL J203802.7$-$400814 & 0.0000 & 0.002 & 0.0000 & 0.002 & 20.26 \\
 & 0.1354 & 0.002 & -2.0863 & 0.002 & 19.65 \\
 & -2.1721 & 0.002 & -0.3813 & 0.002 & 19.90 \\
 & -1.3791 & 0.002 & -2.0573 & 0.002 & 19.61 \\
GraL J210329.0$-$085049 & 0.0000 & 0.007 & 0.0000 & 0.008 & 20.43 \\
 & -0.9107 & 0.038 & 0.0664 & 0.017 & 19.08 \\
 & -0.2587 & 0.002 & 0.7528 & 0.002 & 18.47
\enddata
\label{table:data_for_models}
\end{deluxetable*}

% TABLE 6:  ASTERISMS
% \scriptsize
\begin{deluxetable*}{lccccl}[ht!]
% \tablewidth{0pt}
\tablecaption{Invalidated GraL lens candidates (i.e., asterisms).}
\tablehead{
\colhead{Name (Type)} &
\colhead{Coordinates} &
\colhead{Night} &
\colhead{PA} &
\colhead{Exp. Time (s)} &
\colhead{Notes}}
\startdata
J0001+2040    & 00:01:39,   +20:40:00 & N05-P &  97$\deg$ & $2\times600$ & quasar ($z=1.542$) + star \\
J0009+3719   & 00:09:52,   +37:19:11 & N05-P & 320$\deg$ & $2\times600$ & quasar ($z=0.155$) + star \\
J0015+4853    & 00:15:20,   +48:53:07 & N17-K & 210$\deg$ & $2\times300$ & quasar ($z=0.255$) + star \\
                 &                       & N17-K & 262$\deg$ & $2\times300$ & quasar ($z=0.255$) + star \\
J0036+2751    & 00:36:04,   +27:51:14 & N05-P &  53$\deg$ & $2\times600$ & quasar ($z=1.138$) + star \\
J0101+6719    & 01:01:23,   +67:19:06 & N08-K & 164$\deg$ & $2\times300$ & star + star \\
                 &                       & N08-K &  74$\deg$ & $2\times300$ & star + star \\
J0103+3607    & 01:03:20,   +36:07:05 & N05-P &  48$\deg$ & $2\times600$ & star + likely obscured AGN \\
J0109+5002    & 01:09:46,   +50:02:45 & N17-K & 239$\deg$ &          300 & AGN ($z=0.200$) + star + star \\
J0112+3248   & 01:12:27,   +32:48:38 & N06-K &  15$\deg$ & $2\times300$ & quasar ($z=0.366$) + star \\
J0115+5625    & 01:15:59,   +56:25:06 & N06-K & 356$\deg$ & $2\times300$ & star + star \\
J0218+7445    & 02:18:23,   +74:45:23 & N09-K & 145$\deg$ &          300 & close stellar pair (or single star)
\enddata
\label{table:asterisms}
\tablecomments{The first ten asterisms are presented here to illustrate the format of this table.  The on-line version of the journal contains the full list of invalidated GraL lens candidates.}
\end{deluxetable*}

\clearpage
\end{document}